\documentclass[14pt]{article}
\pdfoutput=1
\usepackage{amsmath,amssymb,amsfonts,amsthm,bm,bbm,cancel,wasysym}
\usepackage{epsfig,graphics,graphicx,epstopdf}
\usepackage{array,booktabs,colortbl,colordvi,multirow}
\usepackage{colordvi,color,xcolor}
\usepackage{hyperref}
\usepackage{rotating}

\usepackage{verbatim}
\usepackage{cite}
\usepackage{subfig}
\usepackage{setspace}
\usepackage{url}
\usepackage[percent]{overpic}
\usepackage{slashed}
\usepackage{authblk}
\usepackage{xspace}
\usepackage{fullpage}
\usepackage{hyperref}

\def\ie{{\it i.e.}}
\def\eg{{\it e.g.}}
\def\etc{{\it etc}}

\def\to{\rightarrow}

\newskip\zatskip \zatskip=0pt plus0pt minus0pt
\def\matth{\mathsurround=0pt}
\def\lsim{\mathrel{\mathpalette\atversim<}}
\def\gsim{\mathrel{\mathpalette\atversim>}}
\def\atversim#1#2{\lower0.7ex\vbox{\baselineskip\zatskip\lineskip\zatskip
  \lineskiplimit 0pt\ialign{$\matth#1\hfil##\hfil$\crcr#2\crcr\sim\crcr}}}

\begin{document}


\begin{flushright}
SLAC-PUB-17178\\
\today
\end{flushright}
\vspace*{5mm}

\renewcommand{\thefootnote}{\fnsymbol{footnote}}
\setcounter{footnote}{1}

\begin{center}

{\Large {\bf Kinetic Mixing, Dark Photons and Extra Dimensions II:  ~Fermionic Dark Matter}}\\

\vspace*{0.75cm}

{\bf Thomas G. Rizzo}~\footnote{rizzo@slac.stanford.edu}

\vspace{0.5cm}

{SLAC National Accelerator Laboratory}\ 
{2575 Sand Hill Rd., Menlo Park, CA, 94025 USA}

\end{center}
\vspace{.5cm}

\begin{abstract}
 
\noindent

Extra dimensions can be very useful tools when constructing new physics models. Previously, we began investigating toy models for the 5-D analog of 
the kinetic mixing/vector portal scenario where the interactions of bulk dark matter with the brane-localized fields of the Standard Model are mediated by a massive $U(1)_D$ dark photon also 
living in the bulk. In that setup, where the dark matter was taken to be a complex scalar, a number of nice features were obtained such as $U(1)_D$ breaking by boundary conditions without the 
introduction of a dark Higgs field, the absence of potentially troublesome SM Higgs-dark singlet mixing, also by boundary conditions, the natural similarity of the dark matter and dark photon 
masses and the decoupling of the heavy gauge Kaluza-Klein states from the Standard Model. In the present paper we extend this approach by examining the more complex cases of Dirac and 
Majorana fermionic dark matter. In particular, we discuss a new mechanism that can occur in 5-D (but not in 4-D) that allows for light Dirac dark matter in the $\sim 100$ MeV mass range, even 
though it has an $s$-wave annihilation into Standard Model fields, by avoiding the strong constraints that arise from both the CMB and 21 cm data. This mechanism makes use of the presence 
of the Kaluza-Klein excitations of the dark photon to extremize the increase in the annihilation cross section usually obtained via resonant enhancement. In the Majorana dark matter case, we 
explore the possibility of a direct $s$-channel dark matter pair-annihilation process producing the observed relic density, due to the general presence of parity-violating dark matter interactions, 
without employing the usual co-annihilation mechanism which is naturally suppressed in this 5-D setup.

\end{abstract}

\renewcommand{\thefootnote}{\arabic{footnote}}
\setcounter{footnote}{0}
\thispagestyle{empty}
\vfill
\newpage
\setcounter{page}{1}



\section{Introduction}

Although its true nature remains in the realm of speculation, the presence of dark matter (DM) clearly signals the existence of new physics beyond the Standard Model (SM). It remains 
unknown whether or not DM interacts other than gravitationally with the SM although models that attempt to calculate the observed DM relic density generally postulate that such interactions 
exist but they are likely to be far weaker that the known weak interactions. Until rather recently, Weakly Interacting Massive Particles (WIMPS)\cite{Arcadi:2017kky} and 
axions\cite{Kawasaki:2013ae,Graham:2015ouw} were the leading contenders for DM as their existence arises from UV-complete frameworks, such as Supersymmetry, or from attempts  
to address other issues such as the strong CP problem. While very important searches for these particles are continuing, the lack of any positive evidence for these scenarios necessitates 
that we widen the scope of potential DM candidates as well as the techniques to search for them\cite{Alexander:2016aln,Battaglieri:2017aum}.  One scenario which is gotten significant recent 
attention is the kinetic mixing/vector portal model\cite{vectorportal,KM} wherein one posits a new dark $U(1)_D$ gauge field, the dark 
photon (DP), as a mediator of the interaction between DM and the SM. This gauge field is the dark sector analog of the photon in ordinary QED except that the corresponding gauge boson is 
generally massive. The DP interaction with the SM fields is generated via the kinetic mixing (KM) of this new gauge field with the SM hypercharge 
$U(1)_Y$ via loops of particles charged under both gauge groups and is characterized by a mixing strength parameter $\epsilon \sim 10^{-(3-4)}$ or so. For DM and DP masses in the 
$\sim 10-1000$ MeV range, this interaction is of sufficient strength to allow the cross section for DM annihilating into SM particles to have the correct magnitude required for the DM to reach its 
observed abundance via the familiar thermal freeze out (FO) mechanism, \ie, the DM here is a thermal relic as in the WIMP scenario. The general parameter space of this model framework is 
currently being explored by multiple existing experiments and will be further examined in great detail by numerous experiments planned for the future that employ various innovative 
techniques\cite{Alexander:2016aln,Battaglieri:2017aum}.

Extra dimensions (ED) are a useful tool for building interesting models of new physics to address outstanding issues\cite{ED}. In our earlier paper\cite{Rizzo:2018ntg}, hereafter referred 
to as I, we examined a toy 5-D version of the KM model assuming a single, flat, extra dimension\cite{morrissey,keith} which could be described as a bounded interval of inverse size 
$R^{-1}\sim 10-1000$ MeV with the SM fields living on one of the brane boundaries as 4-D objects while the DM and DP experienced the full 5-D. In I, in addition to discussing the general 
setup for such an approach, we considered the case where the DM was a complex scalar field, $S$, such that the DM to SM annihilation process was automatically $p$-wave. This easily 
avoids, as in 4-D, the well-known strong constraints on this cross section arising from the CMB at $z\sim 600$\cite{Ade:2015xua,Liu:2016cnk } and the potentially stricter ones arising from 
21 cm measurements at $z\sim 17$\cite{Bowman:2018yin,Slatyer:2016qyl,21club}.  
The $p$-wave nature of the annihilation process allows it to be velocity(squared) suppressed at later times (due to lower temperatures) but still large enough to yield the necessary rate at FO 
to produce the observed DM abundance. In addition to providing new experimental signatures to search for, this framework accomplished several interesting things: ($i$) The lightest field 
in the DP Kaluza-Klein (KK) tower could obtain its mass via boundary conditions (BCs) without the need to introduce a dark SM singlet Higgs field which obtains a vev and spontaneously 
breaks the $U(1)_D$. ($ii$) The DP and DM masses are naturally of the same order $\sim R^{-1}$ without any fine-tuning. ($iii$) The mixing of $S$ with the SM Higgs could be negated in 
some cases via a choice of BCs thus avoiding potentially dangerous exotic Higgs decays; this can only be done by a fine-tuning in 4-D. ($iv$) The higher KK modes of the DP were shown 
to naturally and necessarily decouple from the SM. Further, given the very weak coupling of the dark sector to the SM, we observed that ordinary SM physics is shielded from most of the 
internal dynamics of this (Abelian) dark sector even if it becomes somewhat strongly coupled at high mass scales. 

In the present paper we extend our previous study to the case where the DM is a fermion. In 4-D, in such a situation, only Majorana (or pseudo-Dirac) fermions are allowed as DM since only 
they naturally lead to a $p$-wave annihilation 
or a co-annihilation process. On the other hand,  Dirac DM annihilating via a spin-1 DP mediator is necessarily an $s$-wave, velocity-independent process and so is excluded by the CMB discussion 
above.  As we will see below, however, going to 5-D allows for a new mechanism, occurring through the destructive interference of the multiple DP KK exchanges. This can produce a sufficiently 
large annihilation cross section at FO while the same process is simultaneously highly suppressed at lower temperatures in a manner similar to, but more significantly than, resonant enhancement. 
The generalization of the 4-D case where the DM mass eigenstates are Majorana fields to 5-D will also be demonstrated to bring something new. In 4-D the single DM fermion has purely 
vectorial couplings to the DP as is required by 
gauge invariance in analogy to QED. Once a Majorana mass is generated by symmetry breaking (with the Dirac mass term already present), the DP couples off-diagonally, connecting the two 
Majorana mass eigenstates\cite{4dmaj}. 
In 5-D, starting from the analogous setup, not only do these fermions couple in an off-diagonal manner to the DP (allowing for co-annihilation when the splitting between the mass eigenstates is 
small) but also a diagonal axial-vector coupling term can be generated. This, as we will see, originates from the fact that the DM couplings to the DP KK tower are generally {\it not} vector-like 
as they are in the 4-D model due to parity-violating brane boundary conditions and this allows for a direct 
$p$-wave annihilation process that is effective even when the Majorana mass splitting of the DM with its heavier partner is large.  Both of these fermionic model possibilities can lead to exotic 
signatures at the experiments that are searching for the production of the DP. Of course, we remind the reader that these are only incomplete 
toy models at this point and should be understood as suggestive frameworks for more complete constructs. However, these successes indicate that more realistic versions of the models of the 
type discussed here need to be pursued. 

The outline of this paper is as follows: In Section 2, we first provide a summary of the common 5-D bulk gauge and scalar physics from I that will be required in our subsequent analyses.  
As we will see, the detailed nature of the bulk scalar (beyond its having a vev) is mostly irrelevant to the present setup as it plays no essential role in the DM annihilation or scattering processes. 
We then discuss the case of a SM singlet, 5-D bulk fermion with a Dirac mass as DM (Model 3) and introduce a mechanism, which is the KK generalization of the familiar resonance enhancement 
scenario, which would allow for $s$-channel DM annihilation. The required enhancement of the freeze-out annihilation cross section in comparison to that near $T\sim 0$ is shown to be of 
order $10^4$. The necessary conditions for this mechanism to function properly are discussed, the various input pieces are analyzed in detail and then a scan of the model parameter space 
is performed to identify regions where these conditions 
are satisfied so that the mechanism can be effective. The specific predictions and properties of several benchmark points having the desired properties in the relevant successful parameter region 
are then discussed at some length.  In particular, it is noted that even though the 5-D theory is vector-like, the DM KK tower states generally have parity-violating couplings to the DP gauge KK states 
due to the DM fermion mixed boundary conditions. The DM direct detection cross section in this setup is shown to be quite small but may be 
potentially observable.  In this scenario, experimental DP searches for {\it either} $e^+e^-$ or missing energy final states should observe signals originating from the decays of lightest  two gauge 
KK tower modes. Section 3 contains a discussion of the case of the DM mass eigenstate being a Majorana fermions (Model 4). Here we choose the dark 
charge of the previously introduced bulk scalar such that it can produce a Majorana mass term after SSB which then splits the existing Dirac state into two, generally far from degenerate, Majorana 
states. With the lightest of these being identified as the DM, we find that co-annihilation is generally not effective in this setup due to the naturally large mass splitting. However, due to the previously 
encountered parity-violating fermion couplings, a $p$-wave annihilation channel via the DM axial-vector coupling to the DP is shown to exist, something not found in the analogous 4-D 
`pseudo-Dirac' models with only vectorial couplings. We again introduce a pair of benchmark scenarios to examine the details of this setup. In this model class, the elastic direct detection process 
is shown to be loop suppressed while inelastic scattering is kinematically forbidden due to the previously mentioned large mass splitting.  The production signals for the DP KK states in this scenario 
are shown to be potentially more interesting and complex than in the corresponding Dirac DM case. Section 4 contains a Summary and our Conclusions.


\section{Model 3: Dirac Fermion Dark Matter}

As is well-known, and as discussed in I, data from the CMB and more recently from 21 cm measurements  highly constrain DM annihilation into, \eg, $e^+e^-$ final states in our mass range of 
interest.  These results exclude cross sections even remotely approaching the canonical thermal freeze-out (FO) value required to reproduce the observed relic density\cite{Steigman:2015hda} 
by factors of order $\gsim 10^{2-4}$,  with the stronger (weaker) constraint applying to lighter (heavier) DM masses. As noted in the Introduction, this is a particularly acute problem if the DM 
annihilation is an $s$-wave process as in this case $<\sigma v_{rel}>$ is roughly velocity/temperature independent so that the values at the time of FO,  the CMB ($z\sim 600$), the 21 cm 
measurements ($z\sim 17$) and today ($z=0$) will be essentially identical. In 4-D, in addition to excluding DM masses in excess of that of the DP (which will also lead to an $s$-wave annihilation 
process with a pair of a spin-1 mediators in the final state), the choice of DM being a Dirac fermion is excluded as this annihilation process, occurring via spin-1 DP exchange, is necessarily 
$s$-wave.  At the very upper end of the DM range of interest to us, $\sim 1$ GeV, 
where, \eg, the CMB data alone requires that $<\sigma v_{rel}>_{CMB}/<\sigma v_{rel}>_{FO} \lsim 10^{-2}$, one might be able to evade this constraint by employing a standard resonance 
enhancement mechanism during FO\cite{Feng:2017drg}. The essential idea behind this mechanism is that at FO, the larger thermal velocities of the DM push their center of mass collision energy 
upwards toward the invariant mass associated with the DP resonance peak (provided the DM mass relative to that of the DP is properly chosen) but then falls back to smaller values for the lower 
temperatures during the CMB and later eras.  This mechanism obviously requires a fortuitous strong tuning of the DM and DP masses so that $2m_{DM}$ is not too far below that of $m_{DP}$; this 
is especially so as these two masses are generally uncorrelated and arise from different sources in the 4-D model setup. The 21 cm constraints, taken at face value, would be roughly an 
order of magnitude more restrictive for which the conventional resonance enhancement would be completely inadequate unless the DM is even more massive. Furthermore, it is clear 
that such a mechanism will be insufficient for satisfying the weaker CMB bound {\it alone} for lower DM masses of order 10 MeV or even 100 MeV.  Clearly, if we want to evade these strong constraints 
for Dirac DM in this mass range some more powerful enhancement mechanism must be active. The fact that we are working in 5-D provides for the existence of such a mechanism to which 
we now turn.

\begin{figure}[htbp]
\centerline{\includegraphics[width=5.0in,angle=0]{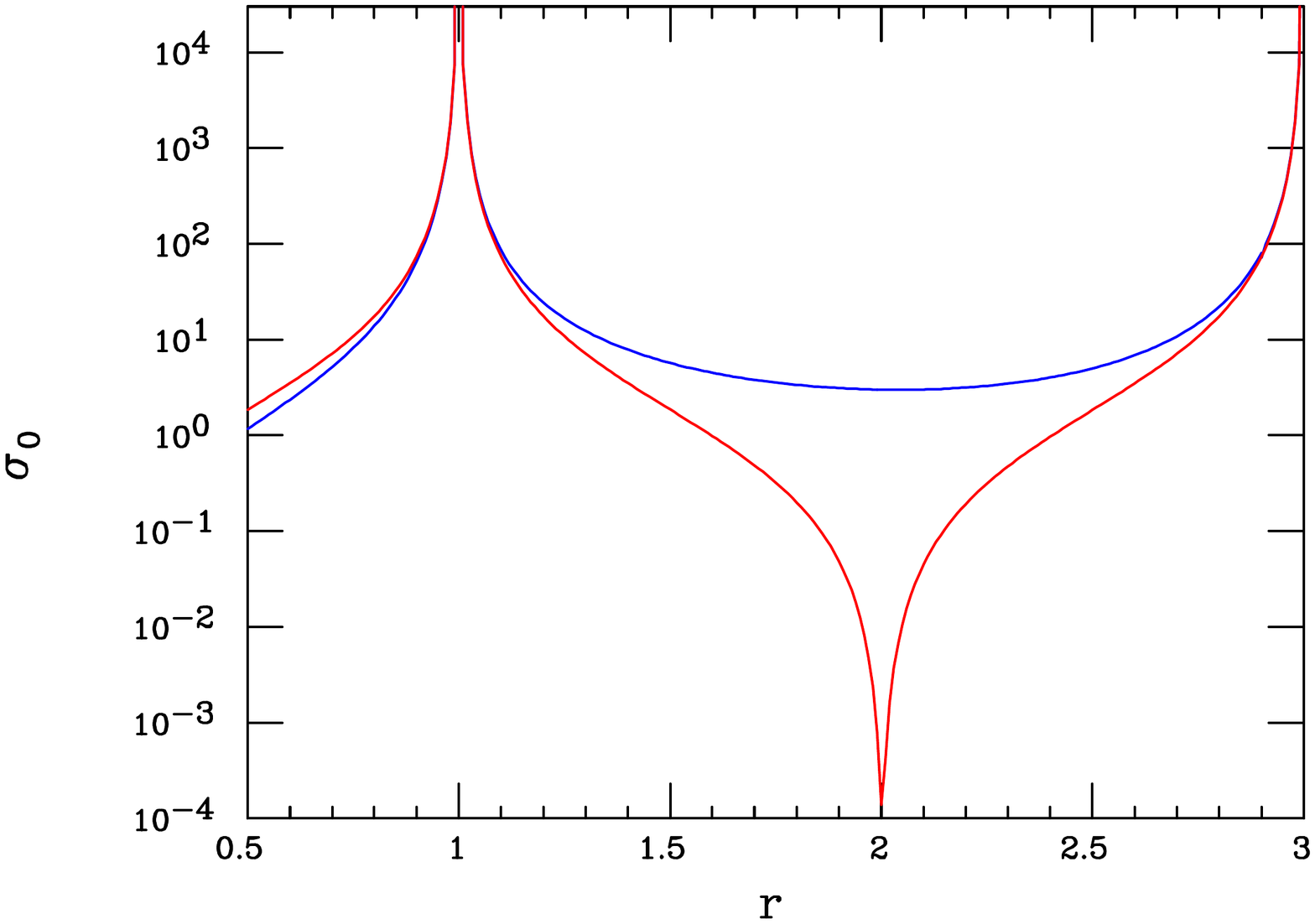}}
\vspace*{-2.5cm}
\centerline{\includegraphics[width=5.0in,angle=0]{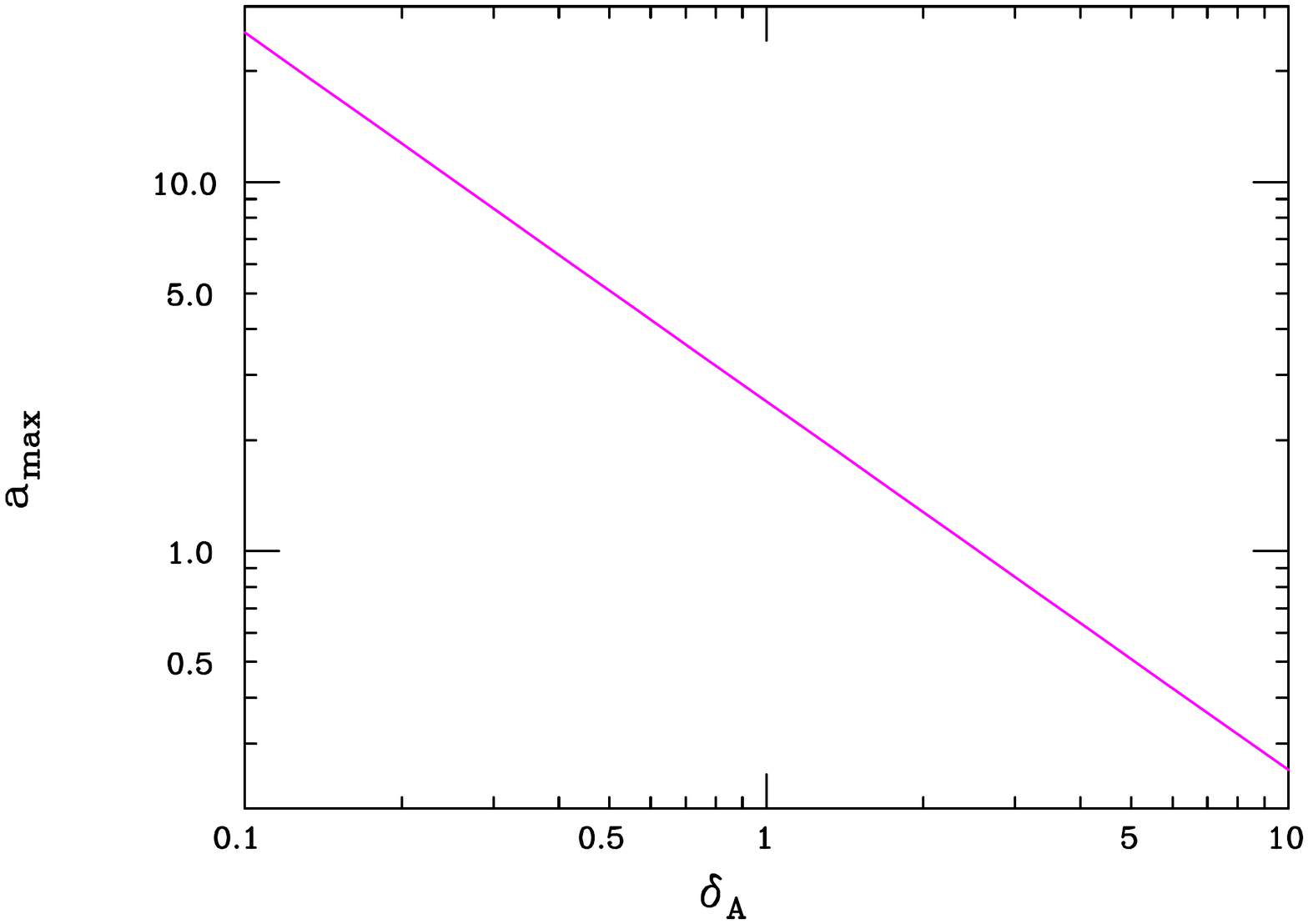}}
\vspace*{-1.30cm}
\caption{(Top) $T=0$ DM annihilation cross section (in arbitrary units) in the simple toy model discussed in the text. The red(blue) curve corresponds to the case of same-sign (alternating sign) couplings for 
the DP KK tower. (Bottom) The upper bound on $a$, \ie, $a_{max}$ as a function of $\delta_A$ in the general vicinity of $\delta_A=1$.}
\label{toy}
\end{figure}

To get the basic idea, consider two very similar versions of a simple toy model for the interaction of Dirac DM with the DP KK tower states assuming for purposes of demonstration 
that the magnitude of the product of the DM/SM couplings are the same for each KK level and that the DP tower masses are given by $M_{V_n}/M_{V_1}=2n-1$, qualitatively similar to the more realistic 
models to be discussed below. Then, again qualitatively, $\sigma_0=\sigma v_{rel}$ for DM pair annihilation into SM final states as a function of $r=\sqrt s/2M_{V_1}\leq 3$ is given, apart from an irrelevant 
overall factor, in the upper panel of Fig.~\ref{toy} when all of the gauge KK tower couplings are identical (red) and where they have the same magnitude but alternate in sign (blue).  Here we observe 
a phenomenon that has been well known since the early days of KK phenomenology\cite{ED}.  When all the KK couplings have the same sign there is a narrow, very strong destructive interference region 
lying between the first two KK resonances, which in fact, in this simple toy example lies exactly halfway between them at $r=2$. (There are analogous destructive interferences between each subsequent 
KK pair.) Note that this strong destructive interference is {\it absent} when the sign of the couplings alternate and $\sigma_0$ behaves `normally' between the first two KK resonances. Further note that 
when this destructive interference is present the ratio of the minimum cross section value between the two KK peaks to that at the top of the second KK peak can be as large as $\sim 10^5$ or even 
greater. This extremely deep and narrow strong destructive interference effect is {\it not} simply the due to the two KK resonances that it lies between but is in fact a {\it collective}  phenomena due to 
the exchange of the entire gauge KK tower and is not obtainable in the 4-D theory, even if a second DP state were to be (somehow) present and have the correct mass and coupling. In our case of 
interest, then, we imagine arranging for 
$\sigma_0$ during the CMB/21 cm (and in the present) era to correspond to that obtained near the destructive minimum while during FO a larger value of $r$ is more representative due to $T\neq 0$ 
effects (and so $\sigma_0$ is significantly larger). In such a case one might be able to evade the Dirac fermion $s$-channel cross section constraints. We will refer to this modified version of the traditional 
resonance enhancement setup as the KK-mechanism.{\footnote {As in the case of ordinary resonant enhancement, a tuning of masses and model parameters will be required. However, in our 5-D model 
this tuning is somewhat {\it less} than in 4-D since the DM and DP masses must necessarily be rather close in magnitude and set by $\sim R^{-1}$, whereas this is not the case in 4-D.}} 

Of course this simple toy model is not applicable in a more realistic situation since as we know from our earlier work that, \eg,  amongst other factors, ($i$) the 
KK gauge masses will not be equally spaced, ($ii$), the relevant couplings of the DM initial state and SM final state both can vary as one ascends the gauge KK tower and will generally vary in 
sign, and that ($iii$) the destructive minimum must occur at smaller values of $r<2$ than in the simple toy model since we still must insure that $m_{DM} < M_{V_1}$ to kinematically forbid DM pair 
annihilation into the $2V_1$ final state. ($iv$) It is also obvious that we need to move the deep destructive interference minima closer to the second resonance peak to allow the $T\neq 0$ affects to push 
the cross section to significantly larger values. Whether or not we can discover a set of model points in our parameter that have the necessary flexibility to achieve these desired goals while satisfying all 
other constraints is a non-trivial question. To address it we must first examine the details of the Dirac DM scenario to assemble the necessary ingredients before we can perform a parameter scan.

Following the familiar lines of our previous 5-D constructions, we consider a setup which has the following components: a flat ED {\it interval} described by a co-ordinate $0\leq y=R \phi  \leq \pi R$ with two 
4-D branes bounding either end and with the SM living on the $y=0$ brane. In the bulk, we have the familiar 5-D $U(1)_D$ gauge field $\hat V^A$ which kinetically mixes with the SM hypercharge 
field, $\hat B^\mu$, on the SM brane and which must have a brane localized kinetic term (BLKT)\cite{blkts}, described by a $\sim$ O(1) dimensionless parameter $\delta_A$, also on the SM brane for the 
reasons described in I. The bulk now also contains a SM singlet, DM fermion field, $X$, which has a 5-D Dirac mass, $m_D$, and that has a dark gauge charge $Q_D(X)=1$ leading to a vector-like 
coupling of $X$ to the DP in the bulk as in the 4-D model. A complex, SM singlet scalar, $S$, is also present in the bulk with $Q_D(S)=Q_D$ and  whose potential leads to a non-zero vev for this 
field, $v_s$. This complex scalar is not required to have a BLKT 
although this addition is straightforward; here, for simplicity, we will assume its absence. Note that a coupling of the form $\bar XXS$ is forbidden 
by gauge invariance so that in the 4-D theory there will be no renormalizable couplings between the Dirac fermion and scalar field towers. The full action for this scenario then takes the form
\begin{equation}
S=S_1+S_2+S_{BLKT}+S_{HS}\,
\end{equation}
where the various pieces are given by 
\begin{equation}
S_1=\int d^4x ~\int_{0}^{\pi R} dy ~\Big[-\frac{1}{4} \hat V_{AB} \hat V^{AB}  ~+\Big(-\frac{1}{4} \hat B_{\mu\nu} \hat B^{\mu\nu} 
+\frac{\epsilon_5}{2c_w} \hat V_{\mu\nu} \hat B^{\mu\nu}  + L_{SM} \Big) ~\delta(y) \Big] \,,  
\end{equation}
which describes the SM plus pure 5-D gauge interaction including the KM on the 4-D brane. The hatted fields must undergo field redefinitions, described in I, to bring this term into canonical form and, as usual,  
$D_A=\partial_A +ig_{5D} Q_D \hat V_A$ is the gauge covariant derivative in obvious notation.
\begin{equation}
S_{2}=\int d^4x ~\int_{0}^{\pi R} dy ~\Big[ i\bar X\Gamma^A D_A X-m_D\bar XX + (D_A S)^\dagger (D^A S) +\mu_S^2 S^\dagger S -\lambda_S (S^\dagger S)^2\Big]\,
\end{equation}
describes the bulk fermion and scalar interactions with $\Gamma_A$ being the 5-D gamma matrices and 
\begin{equation}
S_{BLKT}=\int d^4x \int_{0}^{\pi R} dy ~\Big[-\frac{1}{4}  \hat V_{\mu\nu}  \hat V^{\mu\nu}  \cdot\delta_AR~\delta(y)  \Big]\,
\end{equation}
describes the gauge BLKT term on the SM 4-D brane. Finally, the action also contains the potentially dangerous term 
\begin{equation}
 S_{HS} =\int d^4x ~\int_{0}^{\pi R} dy  ~\lambda_{HS} H^\dagger H S^\dagger S~\delta(y) \,
\end{equation}
with $H$ being the SM Higgs field, which we can either render benign as in I by a choice of BCs or we can simply perform an appropriate fine-tuning of $\lambda_{HS}$ as is done in 4-D; we 
will make the former choice here. 

With such a large compactification radius, $R^{-1} \sim 100$ MeV or so, as we consider here one might have concerns that the dark sector $U(1)_D$ gauge theory may become strongly coupled 
before we reach the $\sim 100$ GeV weak scale relevant for the SM. 
We can use Naive Dimensional Analysis (NDA) to estimate this scale: $\Lambda_{NDA} \sim 16\pi^2/g_{5D}^2 \sim 16\pi^2/ (g_{4D}^2R)$. For a lightest gauge KK mass $m_{V_1}\sim 100$ MeV and 
$g_D=g_{4D}\sim 0.1$, typical of what we will deal with below, one obtains $\Lambda_{NDA}\sim 3.2-3.5$ TeV which is fairly safe as such large mass scales will not be remotely approached in the 
discussions below. Note that this scale corresponds to roughly $N_{KK}\sim \Lambda_{NWA}R \sim 10^4$ KK {\it levels} before the onset of strong coupling. Also, as noted in I, the SM is itself shielded from 
any potential dark sector strong coupling by the tiny sizes of the couplings to the more massive gauge KK states.

As described in I, after field redefinitions the gauge and scalar parts of the present Model 3 are essentially those given by Model 2, except for the interchange of the roles of the $y=0$ and 
$y=\pi R$ branes, which leads to some minor changes, and the fact that the dark charge of $S$ is here not restricted, with the bulk Dirac fermion being essentially the only important new element. 
Given these changes, we briefly summarize some essential aspects of Model 2 with these differences incorporated. 

\begin{figure}[htbp]
\centerline{\includegraphics[width=5.0in,angle=0]{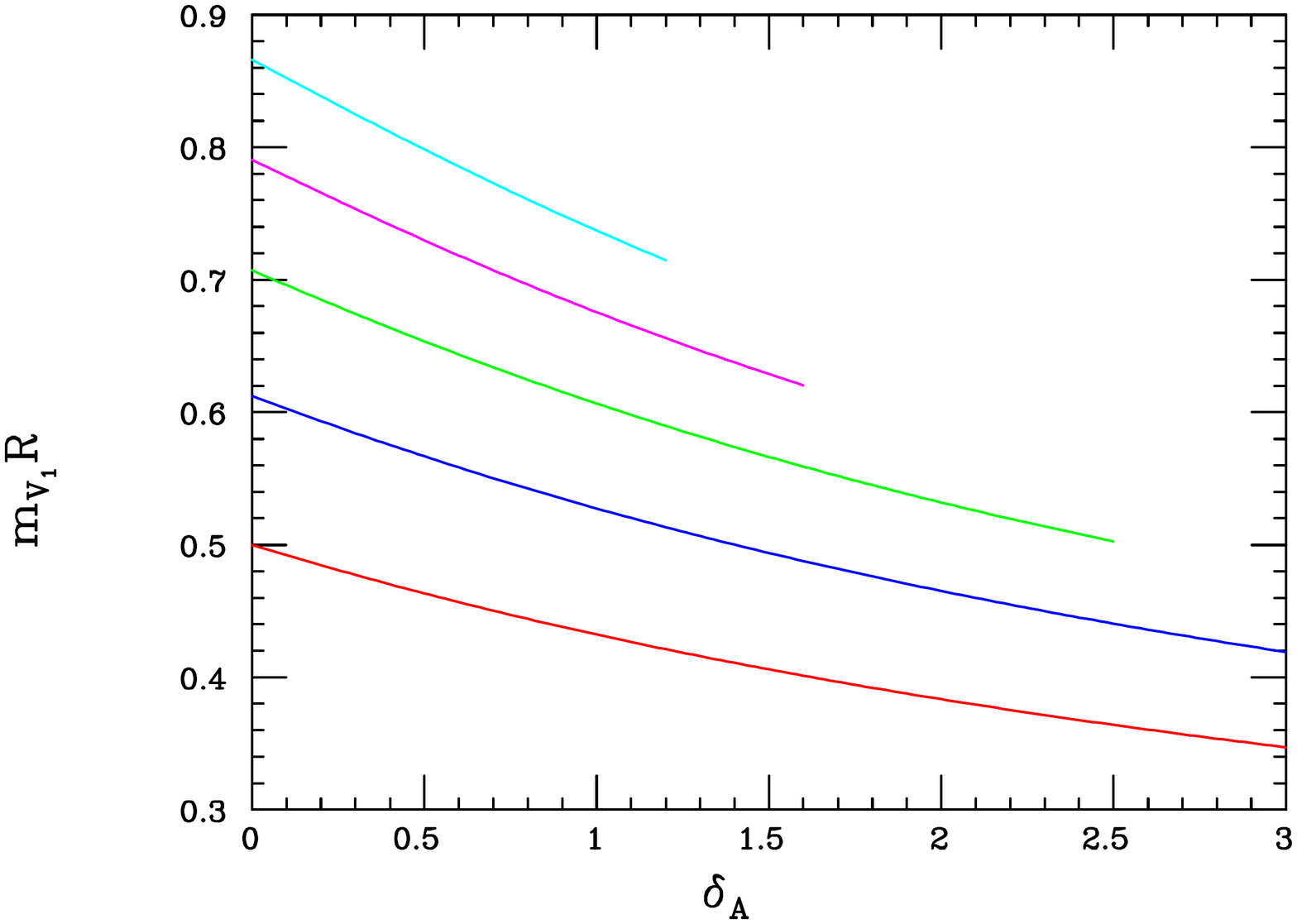}}
\vspace*{-2.5cm}
\centerline{\includegraphics[width=5.0in,angle=0]{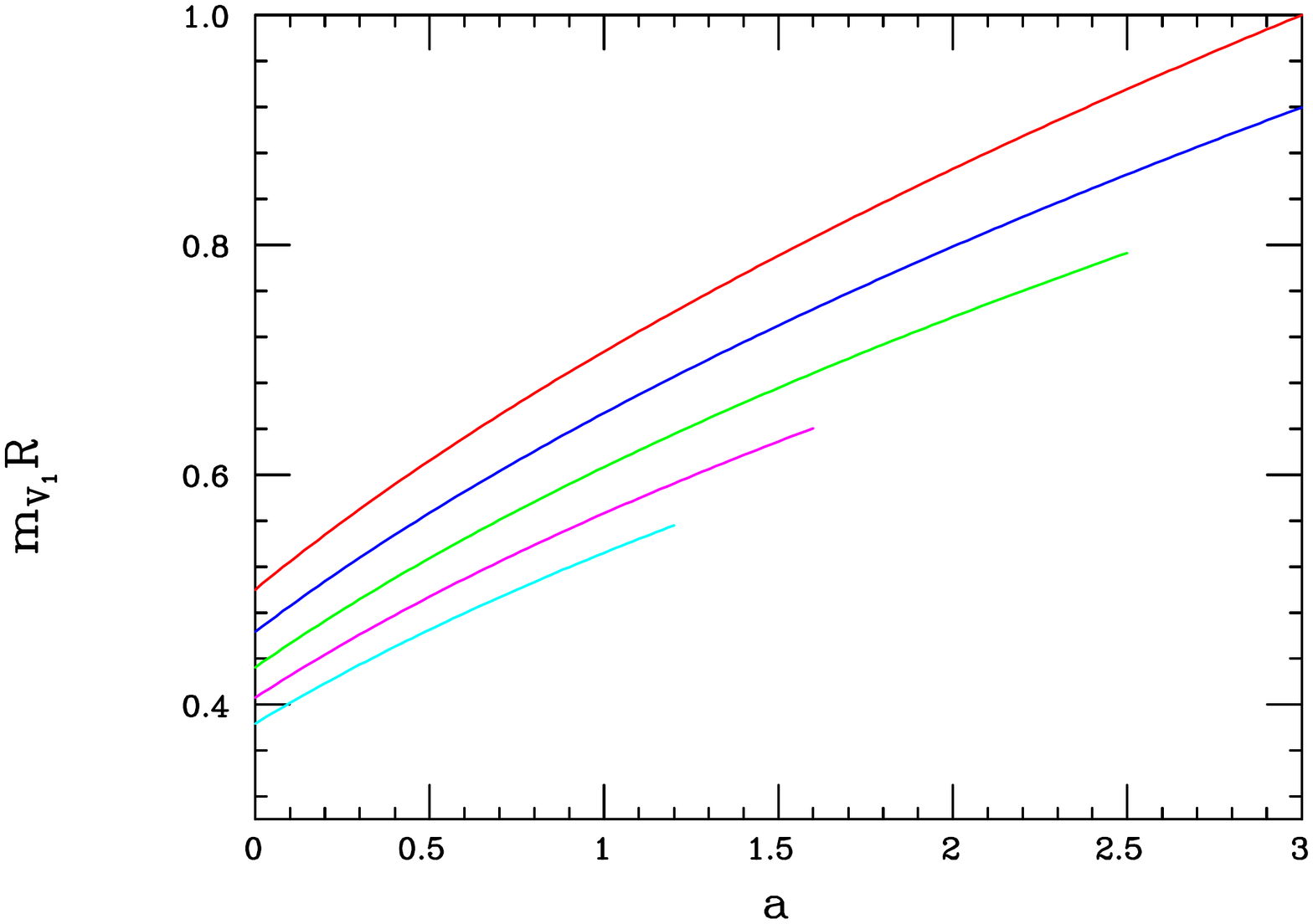}}
\vspace*{-1.30cm}
\caption{ (Top) Lightest gauge KK mass as a function of $\delta_A$ for $a=0(0.5,1,1.5,2)$ corresponding to the red (blue, green, magenta, cyan) curve. (Bottom) Same as above but with the roles of 
$a$ and $\delta_A$ interchanged. Note that that the curves terminate due to the $a_{max}$ bound as discussed in the text.}
\label{res9}
\end{figure}

The vev of the dark Higgs, $S$, produces a bulk mass for the dark gauge field so that masses of the corresponding KK tower fields, $V_n$, are given by 
\begin{equation}
m_{V_n}^2=\Big(\frac{x^V_n}{R}\Big)^2+(g_{5D}Q_Dv_s)^2\,, 
\end{equation}
where the roots $x^V_n$ are found to be given by the solutions of the equation (arising from the BCs) 
\begin{equation}
\cot \pi x^V_n = \frac{\delta_A}{2x^V_n} \Big[ (x^V_n)^2+(g_{5D}Q_Dv_sR)^2\Big]~=\Omega_n\, 
\end{equation}
with the useful dimensionless combination of factors $a=(2g_{D}Q_{D}v_sR)^2$ being of O(1) and frequently appearing in the discussions to follow. For each value of the BLKT parameter, $\delta_A$, 
one finds that there is a  corresponding maximum value for the parameter $a$, $a_{max}=8/(\pi \delta_A)$ (and vice versa) which is shown in the lower panel of Fig.~\ref{toy}. This important 
parameter boundary is easily seen as arising from the root equation above and corresponds to locations where the lowest lying gauge root is being driven to zero. {\footnote {If larger values of $a$ 
were considered, then imaginary values for this lowest root would be obtained although physical, non-tachyonic masses for the lightest gauge mode might still be possible depending upon the 
specific value of $a$ over a narrow range.}} We note 
that this physical boundary will play an important role in the discussion that follows.  Variation of the value of $a$ within its allowed range provides flexibility to adjust the relative contributions of this 
bulk mass term and the `geometric' piece $\sim 1/R$ to the total masses of lowest gauge KK excitations. This balancing of mass sources is necessary since we require the ratio of the masses of 
next-to-lightest to the lightest KK state to be $<2$ for the KK-mechanism to function.  Subject to this constraint the mass of the lightest gauge KK state as a function of both the ($\delta_A,a$) parameters, 
an important consideration in the parameter scan to follow below, is shown in Fig.~\ref{res9}.

The gauge tower 5-D wavefunctions are found to be given by 
\begin{equation}
v_n(y)=N^V_n \big( \cos x^V_n \phi -\frac{1}{t}~\sin x^V_n \phi \big)\,
\end{equation}
where we employ $\phi=y/R$ with $t=\tan \pi x^V_n=\Omega_n^{-1}$ and where the normalization factor is given by
\begin{equation}
(N^V_n)^2 = \frac{2}{\pi R}~\Bigg[1+\Omega_n^2+\frac{\delta_A}{\pi}-\frac{\Omega_n}{\pi x^V_n}\Bigg]^{-1}\,,
\end{equation}
so that the effective KM parameters for the various gauge KK tower states are given by $\epsilon_n=\epsilon_5 N^V_n$ as discussed in I. The important $n$-dependence of these $\epsilon$'s will 
be analyzed below.

Also as discussed in I, the vev of the complex scalar $S$ splits this field into a set of CP-even fields, $h_n$, and a set of CP-odd fields, $\phi_n$, (as $S\to (v_s+h+i\phi)/\sqrt{2}$) which will in general 
mix with the fifth component of the gauge KK fields, $V_{5n}$, to form the Goldstone bosons, $G_n$, and a set of physical CP-odd fields, $a_n$. Actually, since $S$ does not couple to the 
bulk fermion field $X$, as we will discuss below, it plays no important role in our discussion here (outside of it having a vev) and we can be quite agnostic about it as long as we insure that the 
lightest physical $h,a$ KK states are more massive than the lightest fermion tower state (\ie, the DM) which can easily be done by judicious parameter choices. For simplicity and to be definitive, 
we here follow I and employ the results in Ref.\cite{5d}  although we stress that this choice is not necessary for any of the development below. Note that the corresponding BCs in this case remove 
the dark Higgs-SM Higgs mixing at tree-level completely. Following, \eg, \cite{5d}, we then have
\begin{eqnarray}
G_n &=& \frac{\sigma_n V_{5n}+g_{5D}Q_Dv_s \phi_n}{(\sigma_n^2+(g_{5D}Q_Dv_s)^2)^{1/2}} \nonumber\\
a_n &=& \frac{\sigma_n \phi_{n}-g_{5D}Q_Dv_s V_{5n}}{(\sigma_n^2+(g_{5D}Q_Dv_s)^2)^{1/2}}\,,
\end{eqnarray}
where $\sigma_n=(n+1/2)/R$ {\footnote {As in I, we will write these expressions in the form employing KK level dependent mixing angles: $a_n=\cos \theta_n \phi_n-\sin \theta_n V_{5n}$, 
\etc, as will be employed below.}}.  Note that we shift the field $S$ and rewrite the action in terms of $h,\phi$ first before applying any BCs to the solutions of the equations of motion of 
these fields. The Goldstone bosons are, as usual, absent in the unitary gauge in which we will work, while the $a_n$ KK tower fields acquire physical masses that are given by \cite{5d} 
$m_{a_n}^2=(\frac{n+1/2}{R})^2+(g_{5D}Q_Dv_s)^2$. The $h_n$ masses are correspondingly given by $m_{h_n}^2=(\frac{n+1/2}{R})^2+2\lambda_Sv_s^2$; here the 
dimensionless O(1) parameter combination $h=8\lambda_Sv_s^2R^2$ is generally useful and will appear frequently when this sector is discussed. 
As we will see below, unlike in I, we will concentrate on scenarios where the mass hierarchy of the lowest KK modes is given by $m_{\chi_1}< m_{V_1}< m_{h_1} < m_{a_1}$ 
where $m_{\chi_1}$ is the Dirac fermion DM mass. (Of course it is always possible to choose a different mass ordering of the pair of states $a_1$ and $h_1$ without it having any influence 
on the fermion DM phenomenology as discussed above.)  
Note that this choice of mass hierarchy requires that the ratio of parameters $h/a=2\lambda_S/(g_{5D}Q_D)^2 <1$ and that $m_{V_1} < m_{a_1}$ which both are easily 
satisfied over a large part of the 5-D model parameter space. The constraint $m_{V_1}<m_{h_1}$ will place a {\it lower} bound on the ratio $h/a$; since neither 
$h_1$ nor $a_1$ are to be DM, unlike in I, we can permit them to decay rather rapidly. As in I, the effective 4-D couplings among the dark scalars and gauge fields (in units of $Q_D$) are 
determined by the integrals over the products of the 5-D wavefunctions which take the general form (where the $\cos \theta_m$ is the mixing factor defined above) 
\begin{equation} 
g_D \cdot c_{mn}^i = g_{5D} \int_0^{\pi R} ~dy~\cos \theta_m ~a_m(y)h_n(y)v_i(y)\,.
\end{equation}
Note that since there will be both scalar and fermion fields coupling to the gauge KK tower in this model, it is convenient to {\it define} the 4-D gauge coupling here simply in terms of the 
normalization/geometric factors as
\begin{equation}
g_D=g_{5D}~R \Big( \frac{2}{\pi R}\Big)^{3/2}\,.
\end{equation}
In numerical calculations we assume that roughly $g_D \sim 0.1$ as in I. We stress again that these scalars KK states will not play any significant role in what follows outside the existence of the vev 
itself since they do not mediate any important interactions relevant to the DM fermion relic density or direct detection cross section.

Our next step is to determine the Dirac fermion KK mass spectrum and its couplings to the DP KK tower states. We first turn our attention to the equations of motion for the left- and right-handed 
components of the fermion field, $X$. We will denote their corresponding wavefunctions by $f^{L,R}_n(y)$ so that the KK decomposition of $X$ in the action above is given by
 \begin{equation}
 X=\sum_n \Big(P_L f^L_n(y)\chi^L_n(x)+ L\to R\Big)\,,
 \end{equation}
with $P_L$ being the usual helicity projection operator.  We recall that the success of the 5-D fermionic KK decomposition requires, from the intermediate use of integration by-parts in obtaining 
the equations of motion, that these 
wavefunctions must satisfy the coupled BC:  $f^L_nf^R_m(\pi R)-f^L_nf^R_m(0)=0$ for all $n,m$. This type of condition is trivially satisfied in orbifold models but this is not so in the present case. Recall 
that the orbifold BC choice is conventionally employed in 5-D constructions so as to obtain a chiral zero-mode for the relevant, generally SM, fermion(s). This masslessness is not something we desire for 
DM in the present setup hence necessitating a different choice for the  BCs.  Furthermore, just as the requirement of the absence of the Higgs portal above to avoid exotic Higgs decays restricted 
the bulk scalar wavefunction BC on the DM brane, we can follow a similar path with the fermion wavefunctions in order to avoid the presence of a corresponding neutrino portal.  This would take the form 
$\lambda^i_n LH\bar \chi^R_n$, where $L^i$ are the three SM lepton doublets; by requiring that $f^R_n(0)=0$ the $\chi^R$  cannot act as effective RH-neutrinos. We note if we did {\it not} make 
this choice and the set of  $\lambda^i_n$ were to take on a common value then the limit on the invisible width of the Higgs\cite{inv} would require this value to be $<10^{-5}$ since many KK modes of 
$\chi^R$ could contribute to this Higgs decay partial width. By choosing $f^R_n(0)=0$ to avoid this problem, we must also requirethat either $f^{L,R}_n(\pi R)=0$ to fully satisfy the by-parts BC constraint 
above.

\begin{figure}[htbp]
\centerline{\includegraphics[width=5.0in,angle=0]{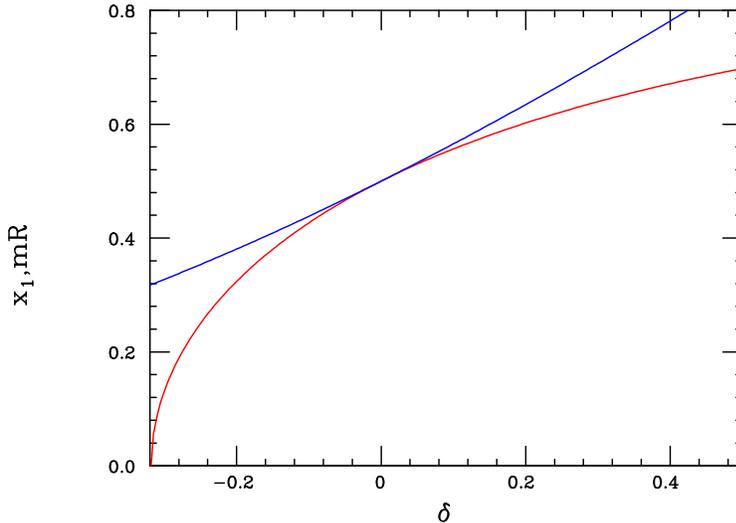}}
\vspace*{-1.90cm}
\caption{The value of the smallest real root (red) and corresponding dimensionless DM mass as functions of $\delta$.}
\label{F-roots}
\end{figure}

Moving forward, we obtain the familiar coupled set of equations of motion of the fermions:
\begin{equation}
(\pm \partial_y -m_D)f^{L,R}_n=-m^F_n f^{R,L}\,,
\end{equation}
where $m^F_n$ are the physical masses of the fermion KK states. If we assume a solution of the form $f^R_n(y)=A_n \cos \sigma y + B_n\sin \sigma y$, the requirement that $f^R_n(0)=0$ 
trivially leads to $A_n=0$. Combining these two first-order equations into a single second-order one also tells us that $\sigma=x^F_n/R$ where the values of $x^F_n$ will be supplied by solving the 
appropriate root equation below so that $(m_n^F)^2=(x_n^F)^2/R^2+m_D^2$. Normalization of the $f^R_n$ wavefunction on the $0\leq y \leq \pi R$ interval further informs us that
\begin{equation}
B_n^2=\frac{2}{\pi R}~\frac{(x_n^F)^2+\delta^2}{(x_n^F)^2+\delta^2+\delta/\pi}\,,
\end{equation}
where we have defined the dimensionless O(1) quantity $\delta=m_DR$. To avoid orbifold BCs and a massless DM zero-mode, we now assume that $f^L_n(\pi R)=0$ to satisfy the integration by-parts 
condition above which then fully determines this set of wavefunctions to be 
\begin{equation}
f^L_n(y)= \sqrt{\frac{2}{\pi R}}~\frac{x^F_n}{\sqrt {(x_n^F)^2+\delta^2+\delta/\pi}} ~\Big(\cos x^F_n \phi+\frac{\delta}{x^F_n}~\sin x^F_n \phi\Big)\,,
\end{equation}
with $\phi=y/R$ as above. This BC choice also leads to the desired root equation
\begin{equation}
x^F_n \cot \pi x^F_n+\delta=0\,,
\end{equation}
which requires that $\delta>-1/\pi$ in order to avoid there being tachyonic roots and/or ghosts in the fermion KK spectrum. Typical values of the smallest root and the corresponding values of the lightest 
fermion (\ie, the DM)  mass as a function of $\delta$ over the interesting range are shown in Fig.~\ref{F-roots}; note that $x^F_1\to 0$ as $\delta \to -1/\pi$ below which value the tachyonic root appears. 
These masses and their $\delta$ sensitivity are important and their variation will necessarily play an important role in the parameter scan to be discussed below. 
Once the purely geometric factors are accounted for as in the scalar case above, the coupling of these left- and right-handed fermion tower fields to the gauge KK tower are given by the integrals 
\begin{equation} 
g_D \cdot g^{L,R~i}_{mn} = g_{5D} \int_0^{\pi R} ~dy ~f^{L,R}_m(y)f^{L,R}_n(y)v_i(y)\,.
\end{equation}
Note that, in all generality, $g^{L~i}_{mn} \neq g^{R~i}_{mn}$ so that the interaction of the DM with the DP tower is now {\it parity-violating} even though the bulk 5-D theory is parity conserving. This 
result is initially puzzling until one recalls 
that this happens to the lowest fermion mode all the time in conventional 5-D orbifold models since there either $f^L_0$ or $f^R_0$ is {\it absent} by construction in order to obtain a massless mode 
and this is seen to be  {\it maximally} parity/charge conjugation violating. Of course in such models the higher fermion modes will have a vector-like coupling but that will not be the case here although 
definite patterns in the KK tower couplings do appear; some details of these couplings will be discussed below. (We remind the reader that here the state $\chi=\chi_1$ 
is to be identified with the DM.) This parity violation as induced by the BCs can lead to the presence of gauge anomalies in the 4-D theory, so that, as is usually done, heavier dark sector fermion 
fields must be introduced into the setup to insure anomaly cancellation but these will play no role in the subsequent discussion.

Summarizing this discussion so far as our future parameter scan is concerned, apart from the overall mass scale set by $R^{-1}$, Model 3 is seen to be described by 4 dimensionless O(1) parameters: 
$\delta_A, a, \delta$ and $h$ which determine all the masses and couplings 
amongst the various KK states. While the gauge KK mass spectrum is controlled by $\delta_A,a$ alone, their couplings to DM is also seen to depend upon $\delta$; the parameter $h$ will play no role 
in what follows. We note that as $a$ becomes larger the lower end of the gauge KK spectrum will become more compressed as the dominant parts of the masses are arising from the Higgs vev and 
not the size of the interval via the KK mechanism. Thus for a given $\delta_A$,  values of $a$ not too far from $a_{max}$ will likely be the most important in obtaining interesting parameter regions 
in our scan below. Similarly, positive but not too large values of $\delta$ yield larger masses for the DM relative to the gauge KK states, as seen in Fig.~\ref{F-roots}, and will likely prove the most important.  

We now can perform the long-awaited scan over the three relevant parameters employing the following chosen scan ranges: $0.1\leq \delta_A\leq 3$, $0 \leq a \leq a_{max}(\delta_A)$, and 
$-1/\pi \leq \delta \leq 1.5$ while simultaneously imposing the requirements that ($i$) the DM mass lies below that of the 
lightest gauge KK state, ($ii$) the product of the SM and DM couplings to the lightest two gauge KK states have the same sign and with the ratio of the DM vector coupling of the second gauge KK 
state to that of the lowest state $\gsim 0.5$ to maximize destructive interference; this also adjusts the important contributions to this destructive interference from all the higher DP KK tower states.
We will also further assume that ($iii$) $0.5m^V_2 \leq 2m_{DM} \leq 0.98m^V_2$ and that $2m_{DM}\geq 1.4m^V_1$. These additional requirements will help push the mass spectra and couplings towards the parameter space regions where the likelihood of the KK-mechanism being the most active  is enhanced and will lead to substantially improved scan efficiency.  The main quantity of 
interest to be first determined during this scan is the DM annihilation cross section during the CMB that we need to be suppressed. Since this is essentially a zero-temperature calculation we can 
rapidly search through the model points in the 
scan to determine whether or not they yield the suppressed values of the annihilation cross section indicative of the strong destructive interference we are seeking. From the set of parameter points 
satisfying this requirement, we extract a large representative set for further study. This will be employed to explore the type and scale of the possible variations in finite temperature cross section 
predictions within this allowed parameter space. Following this, employing these same points, we next determine the ratio, $K$, of their finite temperature 
($x_F=m_{DM}/T=20$ is assumed here) thermally averaged annihilation cross sections at FO to their corresponding values at zero-temperature ($T=0$) keeping only those 
that pass our above criterion. In particular, we will require that $K\gsim 10^4$ for a given point in parameter space to be kept further. 

\begin{figure}[htbp]
\centerline{\includegraphics[width=5.0in,angle=0]{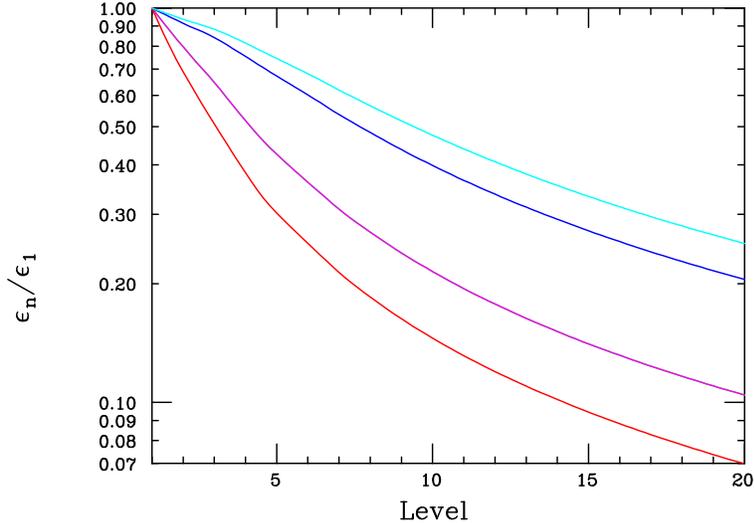}}
\vspace*{-1.30cm}
\caption{Values of $\epsilon_n/\epsilon_1$ appearing in the DM annihilation cross sections as a function of the KK level $n$ for BM1/BM1' (magenta), BM2 (blue), BM3 (red) and BM4 (cyan). The 
anticipated fall-off of this ratio with increasing $n$, assisting the KK summation convergence,  is to be noted.}
\label{eps}
\end{figure}

To perform either of these calculations we must first determine the cross section expression for DM annihilation in this setup. In particular, 
the cross section for $\chi \bar \chi \to e^+e^-$ ($e^+e^-$ being a stand-in here for all of the light, kinematically accessible SM states) in the $m_e \to 0$ limit is given by
\begin{equation}
\sigma=\frac{\alpha g_D^2 \epsilon_1^2}{3\beta_\chi s} ~ \sum_{ij} ~P_{ij} \frac{\epsilon_i \epsilon_j}{\epsilon_1^2} \Bigg[v_iv_j \frac{3-\beta_\chi^2}{2} +a_ia_j\beta_\chi^2\Bigg]\,
\label{sigmae}
\end{equation}
where the sum extends over the intermediate vector KK tower fields, $V_i$. To go further in this calculation, we need to obtain the values of the KM parameter ratios, 
$\epsilon_i/\epsilon_1= N^V_i/N^V_1$, 
which for our final BMs that we will discuss below are easily determined with the results shown in Fig.~\ref{eps}. In this cross section expression $\beta_\chi^2=1-4m_\chi^2/s$, with 
$(v_i,a_i) = \frac{1}{2} (g^{L~ i}_{11}\pm g^{R~ i}_{11})$ being given by the integrals above and 
\begin{equation}
P_{ij}= s^2 ~\frac{(s-m_{V_i}^2)(s-m_{V_j}^2)+ \Gamma_i\Gamma_j m_{V_i}m_{V_j}}{[(s-m_{V_i}^2)^2+(\Gamma_im_{V_i})^2][i ~\to ~j]}\,
\end{equation}
with $\Gamma_i$ being the total width of the KK state $V_i$. All of these quantities are calculable within our set of chosen BMs and are evaluated numerically for the final set of these below. 
For concreteness, the $T=0$ cross section, for which 
$s=4m_{DM}^2$, is generally given numerically by
\begin{equation}
<\sigma v_{rel}>_{T=0} = 2.8\cdot 10^{-30} \rm{cm^3s^{-1}} ~N~\frac{(g_D\epsilon_1/10^{-5})^2}{(m_{V_1}/100 ~\rm MeV)^2}\,
\end{equation}
where $N$ is a roughly O(1) number depending upon the specific parameters of the BM point. Here we see that is easy enough to satisfy both ($T=0$) CMB and 21 cm constraints by, \eg, choosing small 
values of the product $g_D\epsilon_1$. The remaining important issue is whether or not we can obtain parameter points where $K$ is also sufficiently large so as to obtain the observed relic 
density during the higher temperatures at FO. 

\begin{table}
\centering
\begin{tabular}{|l|c|c|c|c|c|c|c|c|c|c|} \hline\hline
    BM  &   $\delta_A$     &   $a$     &   $\delta$  &   $N$  &  $m_{V_1}$ &  $m_{V_2}$ & $m_{DM}$  & Sum  &  $K/10^4$  & $ r_{min}$  \\
\hline
~~1   &     1     &   2.42    &  0.235  &  0.342   & 0.784  & 1.495  & 0.659 &  1.125  & 6.38  & 0.604  \\
~~1'   &    1     &   2.27    &  0.235   &  1.003   & 0.767   & 1.480 &  0.659 & 1.080  &  1.37  & 0.598  \\
~~2    &   0.5     &   3.82    &  0.355  & 0.464   &  1.007  & 1.666  &   0.747 & 1.147 & 4.36   & 0.800  \\
~~3    &   1.5     &   1.61   &   0.180  & 1.518  & 0.642  & 1.380   & 0.620 & 1.027 &  3.20   & 0.402  \\
~~4    &   0.4     &   4.65   &   0.395  & 3.650  & 1.108  & 1.741  & 0.777  & 1.196 &  1.39   & 0.842 \\
\hline\hline
\end{tabular}
\caption{Parameters and general properties of the final five chosen Dirac DM BM models; all masses are given in units of $R^{-1}$. }
\label{spec}
\end{table}

Fortunately, for the wide range of $ 0.4 \lsim \delta_A \lsim 1.5$, we do find parameter points which satisfy all of the above requirements, generally having values $a$ not too far below $a_{max}$ as 
expected. For example, taking $\delta_A=1(1/2,3/2)$ the range of $a$ values which produce `interesting' model points is roughly $2.27(3.82,1.61) \lsim a \lsim 2.54(4.25,1.69)$ with some values clearly 
somewhat preferred over others; this corresponds to roughly a $\sim 5 \%$ parameter tuning somewhat similar in nature what might be required in the 4-D resonance enhancement 
model{\footnote {As noted above, the tuning in the 5-D might be considered {\it less} severe than in the 4-D resonant enhancement model as in 5-D the physical masses of the DM and DP are 
already naturally quite similar.}}.  These ranges are found to produce the required deep, narrow  destructive minima in the $T=0$ cross section which are not far below the position of the second 
DP KK resonant peak. To get a feeling for 
this parameter space, we choose to examine in detail a final set of five representative BM points which have a suggestive spread of parameter values and whose detailed properties are shown in 
Table~\ref{spec}. Note that these are {\it not} best fit points but are selected from those appearing at random in the final selections made by the scan. Amongst other things, we note that all of these 
BM points lead to values $N$ (as defined above) which are not too far from unity as expected and, in particular, produce values of the cross section ratio $K$ in excess of $10^4$ as is required.  

\begin{figure}[htbp]
\centerline{\includegraphics[width=5.0in,angle=0]{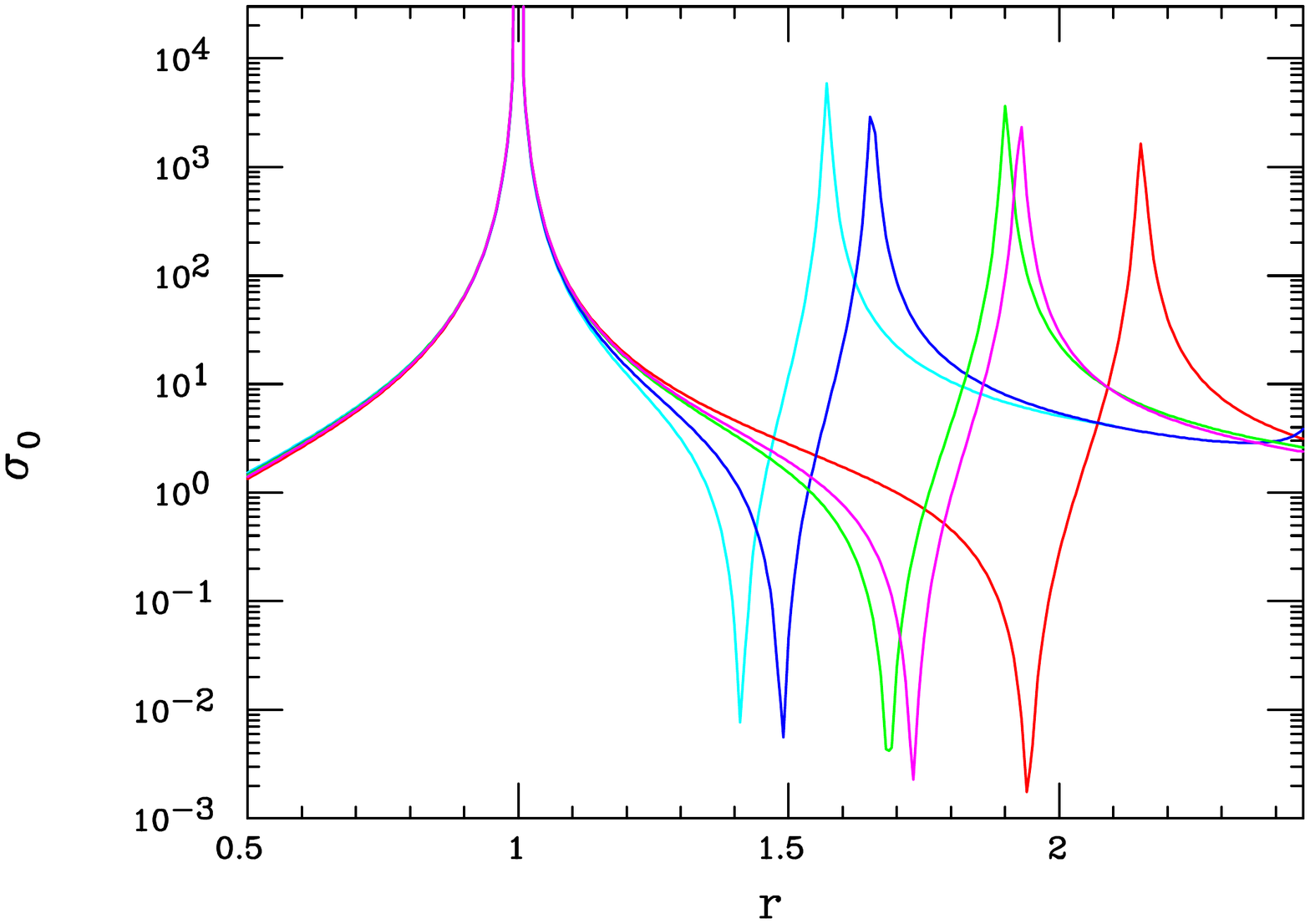}}
\vspace*{-2.5cm}
\centerline{\includegraphics[width=5.0in,angle=0]{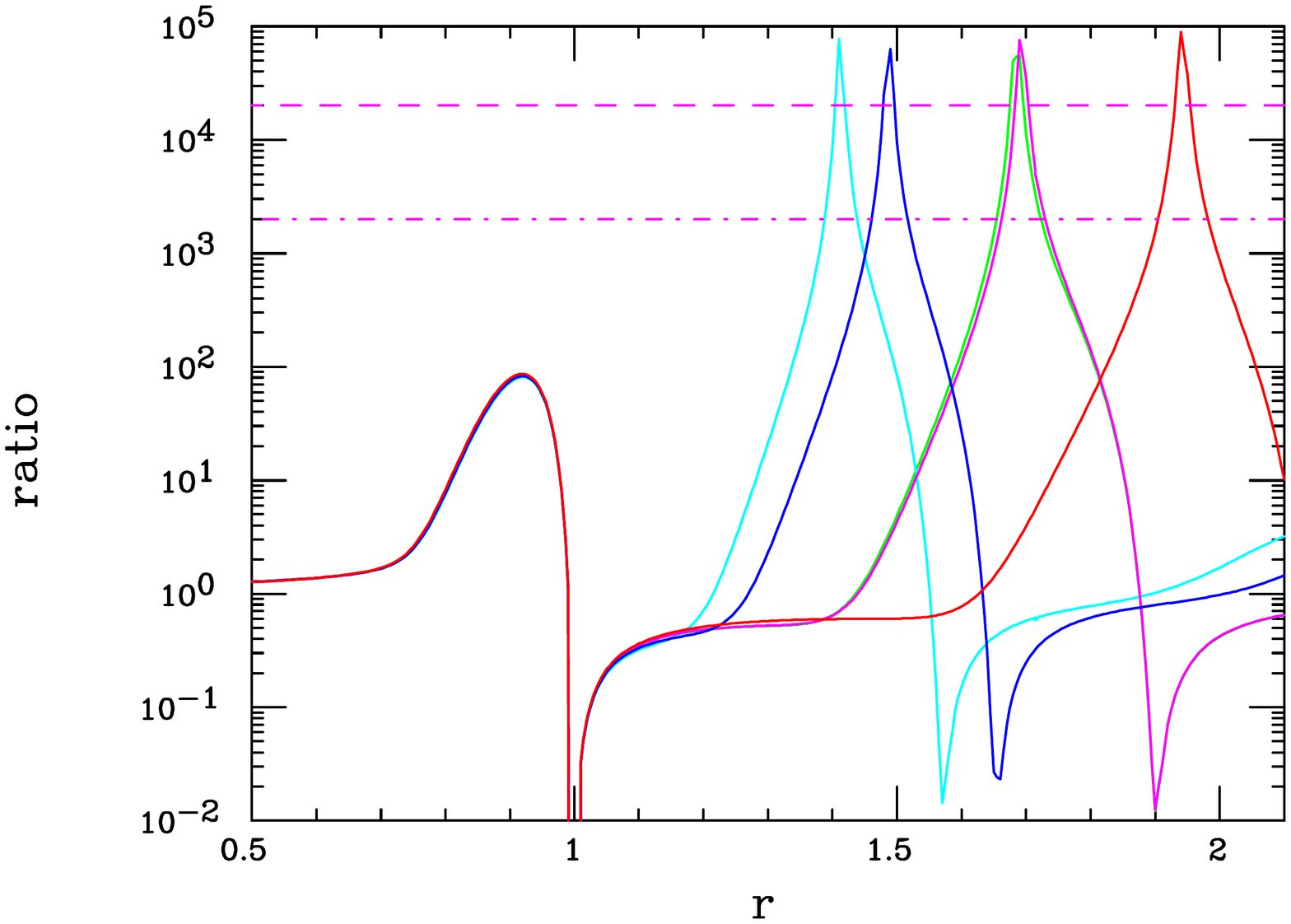}}
\vspace*{-1.30cm}
\caption{ (Top) Scaled $T=0$ DM annihilation cross section as a function of $r=2m_{DM}/m_{V_1}$ for the benchmark models BM1 (green), BM1' (magenta), BM2 (blue), BM3 (red) and BM4 (cyan). 
(Bottom) Ratio of the freeze-out to $T=0$ DM annihilation cross sections, $K$, as a function of $r$ for the same BMs as in the top panel.  The horizontal dashed (dash-dotted) lines are guides to the 
eye for $K=2\cdot 10^4(2\cdot 10^3)$.}
\label{res1}
\end{figure}

We now investigate these BM points a bit more fully.  First, we perform the following instructive exercise: recalling that all of the DM couplings to the KK tower gauge fields are dependent of the 
DM mass itself we {\it freeze} these couplings to their specific values for each BM point 
and explore how the cross sections of interest depend kinematically upon the ratio $r=2m_{DM}/m_{V_1}$; all other parameters for each BM will be held fixed while this analysis is being undertaken. 
The top panel in Fig.~\ref{res1} shows, apart from a common overall factor, the $T=0$ DM annihilation cross section as a function of $r$ for these five BMs. This is the more realistic version of the 
simple toy model result shown earlier in the top panel of Fig.~\ref{toy}.  All BMs show very similar first resonance cross section peaks arising from the lowest KK, $V_1$, and these are analogous to what 
is obtained in the 4-D theory with ordinary resonant enhancement.  However, their detailed behaviors are seen to differ for larger $r$ values due 
to their very different mass spectra (\eg, the mass of $V_2$ which we see as the second resonance peak) and multiple coupling variations. However, all of these BMs are observed to have very strong 
destructive minima in the required range $r<2$, specifically, for $1.41\lsim r \lsim 1.94$. Also we note that the corresponding $V_2$ peak is seen to lie not very far above this deep minima with a typical 
separation of $\Delta r \simeq 0.2$ observed in all cases. This is far smaller than the value of $\Delta r=1$ that was obtained in the simple toy model above. This smaller second KK mass separation is 
a key ingredient for the success of these models since the greater temperatures at FO are somewhat limited as to how much higher they can push the DM collision center of mass energy.


\begin{figure}[htbp]
\centerline{\includegraphics[width=5.0in,angle=0]{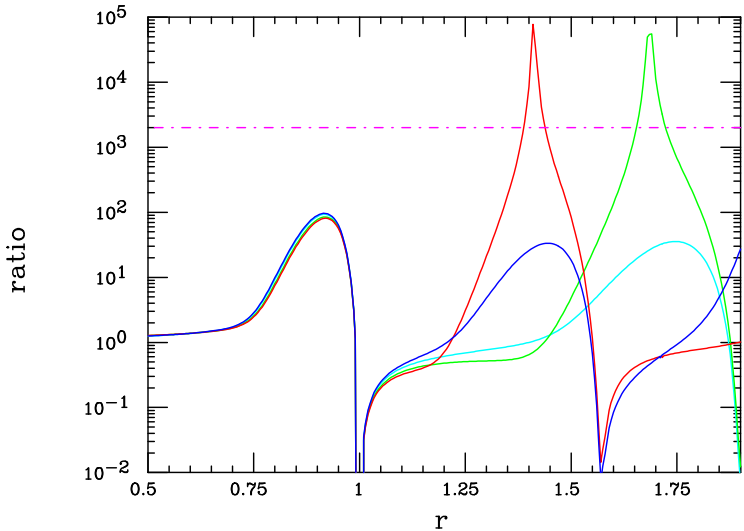}}
\vspace*{-1.30cm}
\caption{Same as in the lower panel of the previous Figure for BM1 (green) and BM4 (red) but also for these same BM (in cyan and blue, respectively) where we modify the signs of the 
product of couplings of the SM and DM of the KK states as discussed in the text for comparison. The horizontal dash-dotted) line is a guide to the eye for $K=2\cdot 10^3$.}
\label{res2}
\end{figure}

The lower panel in Fig~\ref{res1} shows the ratio of DM annihilation cross sections at FO to those at $T=0$, $K$, as a function of $r$, performing the same type of analysis as in the top panel, \ie, 
freezing all other parameters and simply varying $r$. Here we see several things: ($i$) below the $V_1$ peak we observe the conventionally expected resonant enhancement of $K\sim 80-100$ similar 
to that which one obtains in the analogous 4-D models and which is, as discussed above, insufficient for our present purposes. ($ii$) For larger $r$ we see the KK-enhancement peaks associated with the 
deep destructive interference troughs in the top panel.  In all cases, the actual $K$ value at the peak is somewhat {\it larger} than for our randomly chosen BMs indicating that the parameters for these 
representative chosen scan points are not completely optimal in maximizing the possible value of $K$. However, in all cases we see that values of $K \sim 10^4$ are relatively easy to obtain employing 
this mechanism assuming a $\sim 5-10\%$ mass range tuning similar in some qualitative ways to 4-D. To understand the importance of 
the relative signs of the couplings of $V_{1,2}$ (and the rest of the DP KK tower) to the DM, we show in Fig.~\ref{res2} the same results for $K$ as in Fig.~\ref{res1} for BM1 and BM4 but now comparing to 
the same BMs after flipping the sign of the $V_2$ couplings (and which subsequently also modifies the relative signs of the higher KK states). We first see that the normal resonance 
enhancement below $V_1$ is insensitive to this coupling change as it should be. However, when these coupling signs are altered,  we find that the very large $K$ values for our BMs $\sim 10^{4-5}$ are 
now reduced to values of $K \sim 30-40$ that might be obtained by ordinary resonance enhancement. This demonstrates the need for the $V$'s to have the same sign couplings to obtain large $K$ 
values and is exactly what we expected qualitatively from the top panel in Fig.~\ref{toy}. 

For these same interesting BMs, other observables can now be determined as all their parameters are fixed. We can immediately evaluate the $\chi(\bar \chi)-e$ elastic scattering direct detection 
cross section for our BMs, which is given generally (but in the limit of vanishing DM velocities) by 
\begin{equation} 
\sigma_e = \frac{4\alpha \mu^2 g_D^2 \epsilon_1^2}{m_{V_1}^4} ~\Bigg[\sum_i  \frac{\epsilon_i}{\epsilon_1} ~v_i~\frac{m_{V_1}^2}{m_{V_i}^2}\Bigg]^2\,
\end{equation}
where $\mu=m_em_\chi/(m_e+m_\chi)$ is the reduced mass $\simeq m_e$ for the DM masses of interest to us. Note that apart from a common overall factor, all of the BM dependence lies in 
the squared sum within the bracket so that, numerically, we obtain
\begin{equation} 
\sigma_e = 3.0\cdot 10^{-42} {\rm{cm}^2} ~\Big(\frac{100 ~{\rm {MeV}}}{m_{V_1}}\Big)^4 ~\Big(\frac{g_D\epsilon_1}{10^{-5}}\Big)^2  \times \rm{Sum}\,
\end{equation}
where here `Sum' denotes the squared summation above and whose values (also found to be close to unity due to the rapid convergence of the series) for the BM points are given in Table~\ref{spec} 
and which are obtained by employing the results shown in Figs.~\ref{res9} , ~\ref{F-roots} and ~\ref{eps}.  
We note here that the typical values we obtain for this cross section are a factor of order $\sim$ 20-50 below the projected sensitivity of the first full incarnation of, \eg, 
SENSEI \cite{TTYu,Essig:2017kqs,Essig:2015cda}, but may eventually be reached by experiment.

\begin{figure}[htbp]
\centerline{\includegraphics[width=5.0in,angle=0]{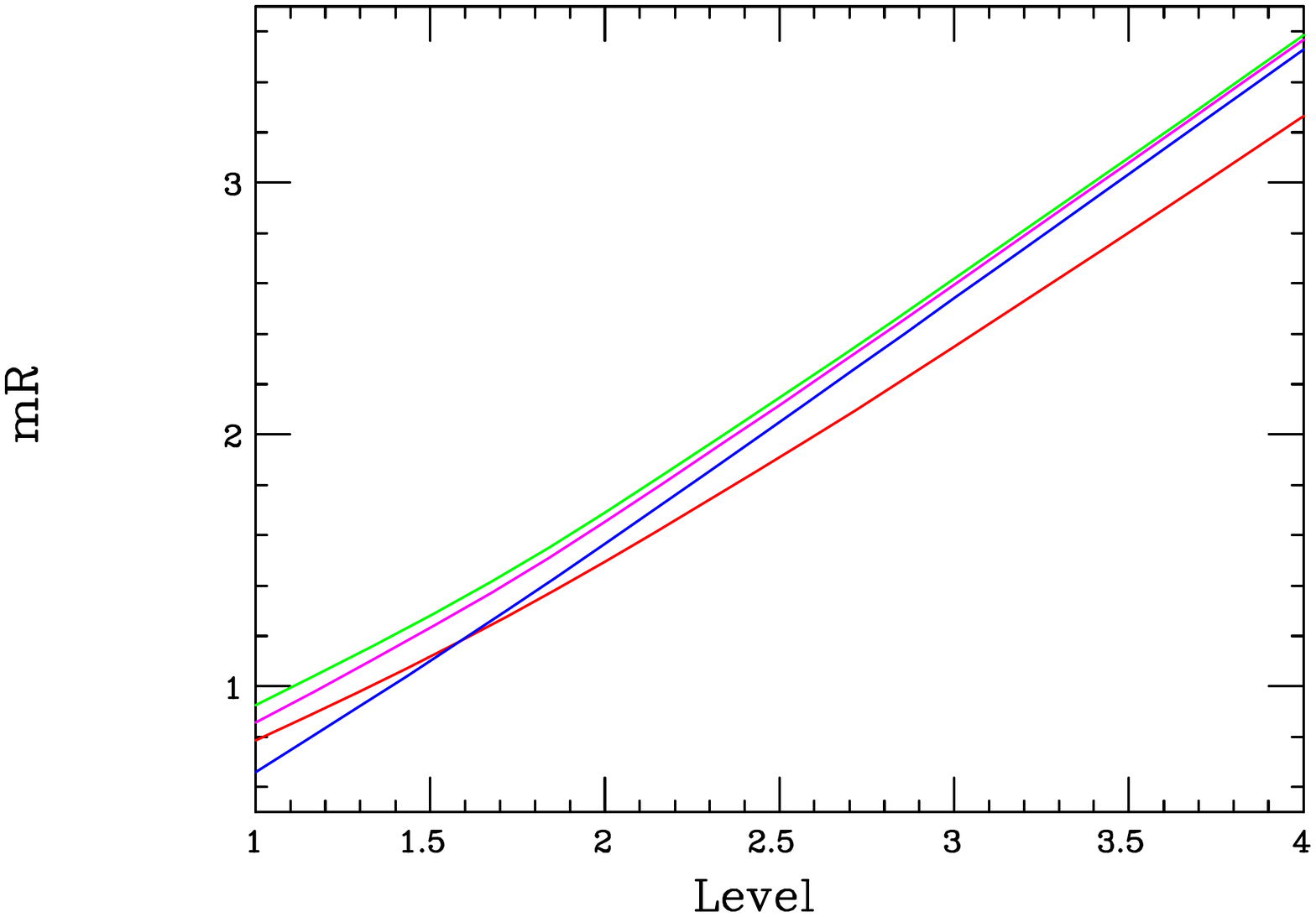}}
\vspace*{-2.5cm}
\centerline{\includegraphics[width=5.0in,angle=0]{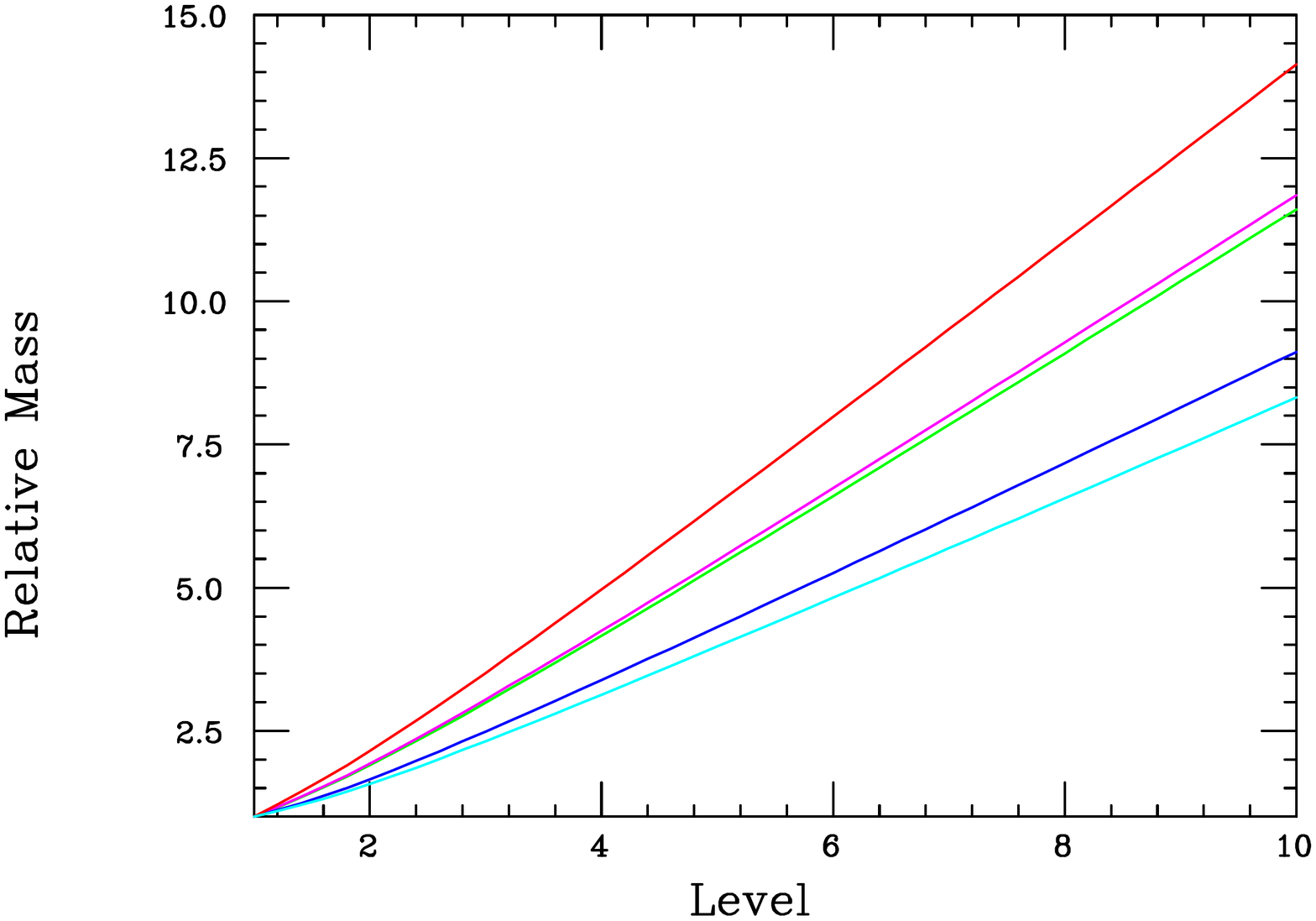}}
\vspace*{-1.30cm}
\caption{(Top) Mass spectrum of BM1 for $r_{ha}=h/a=0.8$ as a function of the level $n$ in units of $R^{-1}$ for  $V_n$ (red), $\chi_n$ (blue), $a_n$ (green) and $h_n$ (magenta), respectively. 
(Bottom) Growth in the mass ratio $m_{V_n}/m_{V_1}$ as a function of $n$ for BM1 (green), BM1' (magenta), BM2 (blue), BM3 (red) and BM4 (cyan).}
\label{sample}
\end{figure}

Turning to the KK mass spectra themselves, the requirements to obtain large $K$ values pushes us into a somewhat constrained location in parameter space which essentially determines the decay 
properties of the lowest lying members of the DP KK tower. Since the DM must be relatively heavy in comparison to $V_1$, this state must {\it necessarily} (unlike in Models 1 and 2)  decay to SM 
particles, \eg, $e^+e^-$ in the 
the mass range of interest to us. The next heavier gauge KK state, $V_2$, is necessarily {\it more} massive than $2m_{DM}$ and since, $g_D^2>>(e\epsilon_2)^2$, essentially only decays to DM 
pairs. Thus $V_1$ is similar in nature to the DP being searched for at HPS\cite{Baltzell:2016eee} while $V_2$ is more like the state decaying to missing momentum/energy which would be sought by 
LDMX\cite{LDMX}.  As will be discussed, the heavier gauge KK states have more complex decay patterns.

We now discuss the other properties of these BMs. Given the values in Table~\ref{spec} the only parameter remaining unspecified (and which played no role so far in the discussion) is $h$ or perhaps 
more usefully the ratio $r_{ha}=h/a$ which determines the relative $ h_n$ and $a_n$ mass spectra.  An important aspect of the scalar sector, as noted above, is that the $h_n,a_n$ do not interact directly 
with the $\chi$'s without mediation by the vectors, $V_n$. Of course there is some significant flexibility in our analysis as the $h_n,a_n$ are essentially decoupled from the considerations above but we 
will employ the spectrum assumptions made earlier for purposes of demonstration; other mass spectra will lead the qualitatively similar results. 
The first aspect to address is the manner in which these $h_n,a_n$ might decay; the possibilities are clear once we recall that the relevant gauge interaction is of the off-diagonal form $haV$ with 
likely the most interesting situation involving the lightest KK mode in each tower. Since $a_1$ is assumed to be the most massive of the lowest KK levels by construction, it decays as 
$a_1\to h_1V_n^*,~V_n^*\to e^+e^-$  and, since by assumption $m_{h_1}>m_{V_1}$, then $h_1\to V_1V_n^*$ occurs. Note that this mass ordering requires a different minimum value of 
$r_{ha}$, \ie, $r_{min}$, for each BM point; these are given in Table~\ref{spec}.  This general type of $h_iV_jV_k$ off-diagonal coupling is generated through the dark Higgs vev since only part of the 
masses of the gauge fields arise from this source and the DP KK mass and dark Higgs coupling matrices are not simultaneously diagonalizable.  The decays of $h_1,a_1$ are thus seen to lead to 
rather complex final states with up to three pairs of $e^+e^-$, 
at least one of which is on-shell (by construction) arising in the case of $h_1$ decay. Since the $a_1-h_1$ splitting can be accidentally small it is possible that the $a_1$ is long-lived in a situation 
parallel to that discussed in I. As the KK towers are ascended one finds that rather complex decay 
patterns can be encountered; note that level-by-level as $n$ increases so does the mass degeneracies between the $\chi_n, a_n$ and $h_n$ states while the $V_n$ generally remain somewhat lighter 
due to the presence of the BLKT.  This large-$n$ behavior is found to be independent of the details of scalar sector unless it too has a BLKT (a possibility which we have ignored here for simplicity). 

A sample spectrum for BM1 is shown in the upper panel of Fig.~\ref{sample} assuming $r_{ha}=h/a=0.8> r_{min}$ for purposes of demonstration; the states are the {\it least} degenerate 
in the case of the lowest tower members. These mass spectra can then be used to determine which decay modes are kinematically open for the various KK tower states. For example, for BM1, we see that 
the following on-shell decays are kinematically accessible: $V_3 \to h_1a_1, \chi_1 \bar \chi_1, \chi_1 \bar \chi_2+$h.c., $a_3\to V_1h_1,V_1h_2,V_2h_1$ and $h_3\to V_1a_1,V_1a_2,V_2a_1$ while 
$\chi_2\to \chi_1V_1$ and $a_2\to V_1h_1$ are the only ones allowed. In this BM case, $h_2$ still only decays off-shell, here to $V_1V_n^*$ and is likely long-lived. For the other BMs the set of allowed 
decays can be somewhat different particularly due to the gauge KK mass spectrum variations which are quite sensitive to choices of $\delta_A,a$. In the lower panel of Fig~\ref{sample} we can 
see how this spacing between the different gauge KK levels, $m_{V_n}/m_{V_1}$, grows with $n$ for the different BMs; since these are the states that are most likely to be produced directly, their spectra 
are of the most immediate relevance. This Figure shows a spread in the masses of almost a factor of $\sim 80\%$ between the different BMs implying that the 
details for the different parameter space points can vary significantly. More generally, we also find the following coupling patterns for the DM tower fields for all the BMs: 
the DM couplings to the KK gauge tower may generally still alternate in sign after the first few gauge levels. More massive fermion KK excitations generally have parity-violating couplings to the DP. 
However, going up the fermion DM tower the couplings eventually begin to alternate between being almost pure vector (axial-vector) couplings, all with the same sign, opposite to that for the DM,  
for odd (even) KK levels making the KK tower couplings, asymptotically, parity-conserving.

\section{Model 4: Majorana Fermion Dark Matter}

The Majorana DM scenario (Model 4) can be obtained by a straightforward augmentation to the action of the Dirac fermion model presented in the last Section;  this is often referred to as the 
`pseudo-Dirac' scenario since both types of mass terms are present simultaneously.  To some extent the Majorana fermion case 
is simpler than the Dirac case since the DM annihilation process will now be automatically $p$-wave if axial-vector couplings to the DP KK states exist or if co-annihilation with its heavier Majorana 
mass partner that arises from the splitting of an originally Dirac fermion into two Majorana components is effective. This means that one is no longer constrained to live near the DM annihilation 
cross section minima when $T=0$ as in the Dirac case above and this leaves significantly greater parameter freedom in the gauge, fermion and scalar sectors. However, it is also simultaneously 
{\it more} complex than the Dirac case since there are now 5 distinct KK towers of fields to deal with although the $h_n,a_n$ fields as above will still play almost no role in our discussion. 

We arrive at this scenario by returning to Model 3 above and choosing the bulk scalar to have $|Q_D|=2$ so that it can generate a Majorana mass term for $X$ when $S$ gets a vev, \ie, 
$m_M=y_Dv_s/\sqrt 2$,  via a new piece in the action given by\cite{4dmaj}:
\begin{equation}
 S_{Maj} =\int d^4x ~\int_{0}^{\pi R} dy  ~ \Big[-y_D  \bar X X^cS +h.c.\Big]\,
\end{equation}
where $y_D$ is a dark Yukawa coupling, a new free parameter which we can simply trade for the Majorana mass itself $m_M=\delta_M/R$. This new additional mass term then alters the 
existing KK equations of motion that we obtained above in the Dirac case to\cite{Huber:2003sf}
\begin{equation}
(\pm \partial_y -m_D)f^{L,R}_n=-(m^F_n -m_M)f^{R,L}\,,
\end{equation}
The solutions to these equations are essentially identical in form to those obtained in the Dirac case as we still impose the same BCs on the solutions: $f_n^R(0)=0$ and $f_n^L(\pi R)=0$ 
although the mass eigenstate structure is now different. In the present case each Dirac mass eigenstate in the KK tower is split into two distinct Majorana KK tower states with the mass values given by 
$m_{1,2n} R=[(x_n^F)^2+\delta^2]^{1/2} \pm \delta_M >0$ where $x_n^F$ and $\delta$ have been defined above. Here, a certain parameter hierarchy and Majorana mass sign convention has been 
assumed, \ie, $\delta_M \geq 0$, so that the physical Majorana tower masses $m_{1,2n}$ are always positive. For example, taking $\delta=\delta_M=0.1$ we find the lowest Dirac KK mass to be 
$\simeq 0.565/R$ from Fig.~\ref{F-roots} and hence the lowest KK values for the two split Majorana states masses $m_{11}$ and $m_{21}$ are $\simeq 0.465/R$ and $0.665/R$, respectively. Note that 
this is quite a sizable mass splitting between these two states, \ie, $(m_{21}-m_{11})/m_{11} \simeq 0.43$ and a strong parameter tuning would obviously be necessary to obtain smaller fractional 
mass splitting values of, \eg, 
a few per cent as would be required for effective co-annihilation.  Given the equations of motion above, the $f_n^{L,R}(y)$ reduce to the same functions as in the Dirac case when written in terms 
of $x_n^F$ and $\delta$ and so the couplings 
$g^{L,R~i}_{mn}$ obtained above are also completely determined once the gauge KK wavefunctions are known. However,  the self-conjugate Majorana mass eigenstates are now just the linear combinations 
$\chi_{1n} =(\chi^L_n +\chi^R_n)/\sqrt 2$ and $\chi_{2n}=i(\chi^R_n-\chi^L_n)\sqrt 2$, respectively{\footnote {It is sometimes convenient to write these in the more conventional $\chi,\chi^c$  
basis\cite{4dmaj}; then $\chi_n=(\chi_{1n}+i\chi_{2n})/{\sqrt 2}i$ and $\chi^c_n=-(\chi_{1n}-i\chi_{2n}){\sqrt 2}i$ as we will generally do here.}}.

Momentarily suppressing Lorentz and KK indices and recalling that, unlike in 4-D, we generally have $g^L\neq g^R$ in the present setup, the interaction of the $\chi$'s with the DP KK tower fields $V$ can 
be symbolically written as $g_D [(g^L\bar \chi  \chi_L+ L \to R) - {\rm{c.c.}}]V$. Thus in terms of the Majorana mass eigenstates $\chi_{1,2}$ this becomes (still suppressing KK indices)\cite{4dmaj} 
\begin{equation}
g_D\Bigg[ \frac{(g^R-g^L)}{2} \Big(\bar \chi_1 \gamma_\mu \gamma_5 \chi_1 + \bar \chi_2 \gamma_\mu \gamma_5 \chi_2 \Big)+\frac{i(g^R+g^L)}{2} \Big(\bar \chi_1 \gamma_\mu \chi_2 - 
\bar \chi_2 \gamma_\mu \chi_1 \Big)\Bigg]V^\mu\,,
\end{equation}
which we see {\it differs} from the 4-D case, where the condition $g^R=g^L$ occurs naturally since the theory is vector-like, with both `diagonal' and `off-diagonal' gauge interactions now being generated.  
In the usual 4-D model case with a single DM field with both a Dirac and Majorana mass term (as noted this is sometimes referred to as the `psuedo-Dirac' scenario) 
we recall that due to gauge invariance, since parity/charge conjugation remains unbroken, only the second term exists due to this purely vectorial nature of the coupling\cite{4dmaj}.  Thus in 
the 4-D case the DM can only reach the desired FO cross section by co-annihilating with its somewhat heavier Majorana partner.  We might expect this reaction to be rather suppressed in the present 
situation due to the typically rather large mass splitting between these two lowest KK states as was encountered above. However, a new direct process now exists and is seen to be $p$-wave 
(hence $v^2$) suppressed due to the axial-vector nature of the relevant coupling (which is present due to the parity violation absent in the 4-D DP model); this is exactly what is required in order to 
satisfy the CMB and 21 cm constraints. The annihilation DM cross section into $e^+e^-$ in the present model is thus of the same form as given in Eq.(\ref{sigmae}) above but with the $v_i \to 0$ and 
an additional overall factor of 2 due to the Majorana nature of the DM, \ie, $a_i^2 \to \frac{1}{2}(2a_i)^2$.  Thus it is easy to numerically obtain the desired relic density with properly chosen values 
of $\epsilon$ and $m_{V_1}$. 

This pair of gauge interactions also has immediate implications for the direct detection searches for DM:  in 4-D at tree-level only the inelastic process, \eg, $\chi_1 e\to \chi_2 e$ is possible and, noting 
that the two masses must be reasonably degenerate in order to obtain the observed relic density in this case, there is a chance that this process might be kinematically open. The corresponding elastic 
scattering process $\chi_1 e\to \chi_1 e$ also {\it does} occur but only at the 1-loop level in the limit of $v_{DM}^2\to 0$.  In 5-D we observe that the inelastic scattering process is irrelevant due to the 
generally large mass splittings that are expected while the new axial-vector coupling of the DM to the DP gauge tower now allows for tree-level elastic scattering. Unfortunately, this is found to be 
$v^2$ suppressed making it likely unobservable anytime soon. Note that the calculation of the couplings $g^{L,R~i}_{mn}$ follows the same path as in the case of Dirac fermions and these remain 
only dependent upon the same three parameters $\delta_A,a$ and $\delta$ as above so that these couplings are independent of the Majorana mass parameter $\delta_M$ which only influences the 
Majorana mass spectrum itself.

Unlike in the Dirac case, however, the presence of the Majorana mass term in the above action generates an interaction between the $\chi_{1,2}$ KK states and those arising from the decomposition 
of the scalar $S$, \ie, $h,\phi$ with the $\phi$ generally mixing as described above with $V^5$ to form the physical state $a$ and the Goldstone bosons (which we ignore). Thus $S_{Maj}$ written 
in the $h,\phi$ basis (but with KK indices remaining suppressed) is simply
\begin{equation}
 S_{Maj} =\int d^4x ~\int_{0}^{\pi R} dy ~  -\frac{m_M}{v_s}\Bigg[\Big(\bar \chi_2 \chi_2-\bar \chi_1 \chi_1 \Big)(v_s+h) + i\Big(\bar \chi_2 \gamma_5 \chi_2-\bar \chi_1 \gamma_5 \chi_1 \Big)\phi \Bigg] \,.
\end{equation}
In 4-D, an interaction of this type could possibly allow for DM annihilation to SM fields via $h$ mixing with the SM Higgs, $H$, but this is absent here due to the BCs for the dark Higgs scalar 
which were assumed above{\footnote {As was noted in the previous Section, the KK modes of $V^5$ do not couple to the SM brane-localized fields.}}. Of course, for other possible choices of these  
scalar BCs we would need to tune $\lambda_{HS}$ to a tiny value which we can always do as is always done in 4-D. Here, above the lowest KK modes, $S_{Maj}$ can provide new decay paths for the  
heavier KK states as we will return to shortly. Also, in 4-D, the fields $\phi$ is absent from the physical spectrum as it plays the role of the Goldstone boson (since $V^5$ is also absent there) 
while here the general combination, $a$, survives as a physical field. However, outside of the presence of the vev, as we have seen, the $h,a$ fields play no important role as far as the dynamics of 
DM is concerned as they either do not couple to the SM or they have couplings which are very highly suppressed depending on the choice of BCs. For the lightest KK levels if we imagine that, 
as previously, $m_{V_1}<m_{h_1},m_{a_1}$, then  $h_1,a_1$ can decay as discussed in the previous Section as this can always be arranged by judicious parameter choices. We still must require that 
$m_{\chi_{1,1}} < m_{V_1} <2 m_{\chi_{1,1}}$ to prevent the $s$-wave DM annihilation into $V_1$ pairs and this implies that the lightest gauge mode is still found to decay to the SM, \ie, $V_1\to e^+e^-$. 
On the other hand, $V_2$ will have a large (possibly dominant) branching fraction into DM pairs, similar to what we obtained above in the Dirac case with the obvious implications for the experiments 
hoping to produce the DP directly. However, the $V_2$ decay to DM plus its heavier Majorana partner is now also possible, depending upon the exact details of the spectra, and this process is seen 
to be kinematically open for our chosen Majorana BM points to be discussed below. In such cases, $V_2$ decay will yield an $e^+e^-$ pair as well as missing energy, something quite different than 
than what is observed in 4-D models. Note that there remains some freedom in the relative position of $\chi_{2,1}$ in the overall mass spectrum. However, we observe that its only allowed decay path 
(since the scalar couplings are `diagonal') is via $\chi_{2,1} \to \chi_{1,1} V_1$.  In such a case it is likely that the $V_1$ will be off-shell (unless $\delta_M$ is of sufficient magnitude) so that several 
of these lowest mass KK states can be long-lived.  Both Fig.~\ref{F-roots} and Fig.~\ref{res2} can be employed to address this issue as the rather strong requirement for an on-shell $V_1$ decay in this 
situation is simply that $2\delta_M> m_{V_1}R$. Clearly in such cases as the Dirac mass parameter $\delta$ increases, the value of $\delta_M$ will also need to increase to insure that the DM mass lies 
below that of $V_1$ in order to avoid s-wave DM annihilation.

\begin{table}
\centering
\begin{tabular}{|l|c|c|c|c|c|c|c|c|c|c|c|c|} \hline\hline
 BM  &$\delta_A$  &   $a$     &   $h$  &   $\delta$  & $\delta_M$ & $m_{V_1}$ & $m_{\chi_{1,1}}$ & $m_{\chi_{2,1}}$  & $m_{h_1}$ &  $m_{a_1}$  & $m_{V_2}$ \\
\hline
~~1 & ~1   & ~1 &  0.9  & 0.1 &  0.1  & 0.607  & 0.465  & 0.665  & 0.673  & 0.707 & 1.389 \\
~~2 &  0.5 & 0.5 & 0.4  & 0.2        & 0.15 &   0.567 & 0.484 & 0.784 & 0.602  & 0.612  & 1.432\\ 
\hline\hline
\end{tabular}
\caption{Parameters of the two Majorana DM BM models as discussed in the text; all masses are given in units of $R^{-1}$. }
\label{spec2}
\end{table}

In order to provide some specific concrete examples of these types of Majorana DM models, we consider the following two BM points, shown in Table~\ref{spec2}, which are somewhat typical, 
random locations in this rather large parameter space yet are qualitatively phenomenologically similar. In units of $R^{-1}$, the masses of the lowest KK excitations in these particular BM cases are 
also displayed in the Table. Note that there is nothing particularly special about these points other than they satisfy 
the rather loose constraints discussed above with many other choices yielding comparable results and leading to similar conclusions, (However, a detailed examination of the full five-dimensional parameter 
space would undoubtably be a useful exercise but this is beyond the scope of the current study.) For these BMs the fractional mass splitting between the two lightest Majorana states is seen 
to be sufficiently large so as to make the co-annihilation mechanism ineffective. One observes that $2m_{DM}$ lies safely away from both $m_{V_{1,2}}$, essentially corresponding to the beginning of the 
relatively flat cross section region between these two resonances similar to points along the blue curve shown in the upper panel of Fig.~\ref{toy}. Thus, if we ignore any co-annihilation contributions due to 
the large Majorana mass splitting as well as any potentially very highly suppressed $h$ exchange terms,  we may safely use the familiar expansion $<\sigma v_{rel}>=bv_{rel}^2$ as well as the machinery 
for the calculation of the annihilation cross section in the previous Section as the DM annihilation process is dominantly $s$-channel via $V_n$ exchange.  This then yields the numerical results (again 
assuming that $x=m_{DM}/T=20$ as above)
\begin{equation}
<\sigma v_{rel}> = 1.2(0.74)\cdot 10^{-25} \rm{cm^3s^{-1}} ~\frac{(g_D\epsilon_1)^2/10^{-7}}{(m_{V_1}/100 ~\rm MeV)^2}\,
\end{equation}
for BM1(BM2) which are not far from the value needed to obtain the observed DM relic density which can be very easily obtained in both cases by suitable parameter choices. Other possible 
BM points will lead to similar results up to similar $O(1)$ factors due to the variations in the fermion and gauge mass spectra and the corresponding KK effective couplings.

\section{Summary and Conclusions}

In this paper we have considered an extension of our previous study of the 5-D kinetic mixing/vector portal model to the case where the DM is fermionic. In these setups the SM singlet bulk Higgs 
sector plays very little role beyond its having a vev.  Unlike in the previously examined scenario of complex scalar DM, the annihilation process for Dirac fermions via the spin-1 DP mediator KK tower 
is necessarily $s$-wave and so is not considered in the 4-D models of this type due to CMB and 21 cm constraints. However, in 5-D, we have discovered  
a new mechanism which allows for this possibility which is the KK generalization of the usual resonance enhancement picture. In that setup, the mass difference $M_{res}-2m_{DM}$ is sufficiently 
small so that the thermal motion of the DM near freeze-out is great enough as to push $\sqrt s \sim M_{res}$ which greatly increases the annihilation cross section at freeze-out.  Thus the 
annihilation cross section can then be smaller by a factor of $K \sim 100$ when $T\sim 0$. This factor is insufficient in the case of the Dirac DM in the mass range of interest to us as the ratio of 
the thermal FO cross section and the upper limit from the CMB and 21 cm data is required to be of order $K \sim 10^4$. In addition, in 4-D the DM and DP masses are not necessarily related so that 
a significant parameter tuning may be necessary for this mechanism to function. To attain such a very large cross section ratio in 5-D we need not only a significant enhancement on 
resonance but also a strong destructive interference below, but not too far away from, the resonance where the $T\sim 0$ DM annihilation takes place. The effectiveness of this arrangement places 
significant requirements on the DM mass and the DP KK tower spectra as well as the SM/DM-DP couplings which must be of the same sign and of similar magnitude for multiple gauge KK levels. 
The depth of the strong destructive interference then results from the well-known collective contributions of the entire DP KK tower. Furthermore, since the DP and DM masses are naturally very similar 
in this 5-D setup one can argue that the parameter tuning is somewhat less in this case than in the analogous 4-D model. Fortunately, in the setup we consider, the cross 
sections of interest were found to be controlled by only 3 parameters. A scan of this parameter space was performed and successful regions satisfying our requirements were identified resulting in the 
5 sample benchmark models that we then examined in greater detail. These BMs were found to yield $K$ values in the range $1.4 \leq K/10^4 \leq 6.4$ and it was further shown that even larger values 
could be obtained with some choices of the parameters. All of these points led to very similar predictions for the DM-electron direct detection scattering cross section.  In such a setup, DP production signals 
in both the $e^+e^-$ and missing energy/momentum channels should be expected from the decays of the lightest two gauge KK states while more exotic decay signatures were possible from the other 
higher KK states.

When the DM is a Majorana field in 4-D, it achieves the required relic density via co-annihilation with its somewhat more massive Majorana partner. This mass splitting is achieved via a DM coupling to 
a SM singlet scalar whose dark charge is chosen so that it can generate a Majorana mass term while a Dirac mass term for the DM is already present, the so-called `pseudo-Dirac' scenario.  
Since these masses are not necessarily related 
this mass splitting can be made small enough so that co-annihilation mechanism is effective. In 5-D, the Dirac and Majorana masses of the DM state are naturally similar so that a much larger mass splitting 
is generated thus negating the efficiency of the co-annihilation process. Fortunately, unlike in 4-D, in 5-D the original Dirac fermion naturally obtains parity-violating couplings to the DP KK tower states which 
are inherited by the physical Majorana fields. This leads to a new coupling of the DM to the DP which is diagonal in the mass eigenstate basis as well as being axial-vector in nature thus leading to a DM 
$p$-wave annihilation process which is needed to avoid the CMB and 21 cm constraints yet can lead to the observed relic density. Due to unbroken parity symmetry such a term is not generated in the 4-D 
DP model when both types of mass terms are present. Unfortunately, this axial vector coupling in 5-D leads to a DM scattering cross section off of electrons which is $v^2$ suppressed so that direct detection 
is unlikely. DP direct production signals in the 5-D case, on the other hand, can be more complex than in the Dirac scenario, even for just the first two DP KK states and can provide unique signatures for this 
scenario at HPS/LDMX-like experiments.  

The extension of the 4-D KM/vector portal to 5-D opens several new windows of opportunity for model building and needs to be examined in more detail than the preliminary studies we have made here.

\section*{Acknowledgements}
The author would like to particularly thank J.L. Hewett for very valuable discussions related to this work. This work was supported by the Department of Energy, Contract DE-AC02-76SF00515.



\end{document}